\documentclass[aps,pra,twocolumn,superscriptaddress,eps2pdf]{revtex4-1}

\usepackage{amssymb}
\usepackage{blkarray}
 \usepackage{amsmath}
 \usepackage{color}
\usepackage{psfrag}
\usepackage{ifpdf}
\ifpdf
\usepackage{epstopdf}  
\fi
\usepackage[raggedright]{titlesec}
\usepackage{multirow}
\usepackage{graphicx}
\usepackage{dcolumn}
\usepackage{verbatim}\usepackage{bm}
\usepackage{algcompatible}
\usepackage{newfloat}
\usepackage{mathtools}

\newcommand{\svdots}{\raisebox{3pt}{$\scalebox{.75}{\vdots}$}}

\DeclareFloatingEnvironment[
    fileext=loa,
    listname=List of Algorithms,
    name=ALGORITHM,
    placement=tbhp,
]{algorithm}

\usepackage{tikz}
\usetikzlibrary{shapes.geometric, arrows}
\tikzstyle{startstop} = [rectangle, rounded corners, text width=1.5cm,  minimum width=1.5cm, minimum height=.5cm,text centered, draw=black, fill=red!30]
\tikzstyle{io} = [rectangle, minimum width=3cm, minimum height=1cm, text centered, text width=3.5cm, draw=black, fill=blue!30]

\tikzstyle{decision} = [diamond,aspect=2.5, minimum width=1cm, minimum height=1cm,text width=3.5cm,, text centered, draw=black, fill=green!30]
\tikzstyle{arrow} = [thick,->,>=stealth]

\tikzstyle{process} = [rectangle, minimum width=3cm, minimum height=1cm, text centered, text width=3.5cm, draw=black, fill=orange!30]

 \newcommand{\ket}[1]{\ensuremath{\vert#1\rangle}}

\newcommand{\kb}[2]{\ensuremath{\vert #1 \rangle \langle #2 \vert}}
\newtheorem{theorem}{Theorem}
\newtheorem{defin}{Definition}

\renewcommand{\Pr}{\ensuremath{\mathrm{Pr}}}
\newcommand{\ps}[1]{\ensuremath{\mathrm{P}_{\mathrm{\mathrm{suc}},#1}}}
\newcommand{\ff}{\ensuremath{\eta}}
\newcommand{\PR}{\ensuremath{\mathrm{PR}}}

\newcommand*{\etc}[1]{\textcolor{black}{#1}}

\def\id{{\mathchoice {\rm 1\mskip-4mu l} {\rm 1\mskip-4mu l} {\rm 1\mskip-4.5mu l} {\rm 1\mskip-5mu l}}}

\begin{document}

\title{Quantum computation with realistic magic state factories}

\author{Joe O'Gorman}
 \affiliation{Department of Materials, University of Oxford, Oxford, OX1 3PH, United Kingdom.}
\author{Earl T.\ Campbell}
 \affiliation{Department of Physics \& Astronomy, University of Sheffield, Sheffield, S3 7RH, United Kingdom.}

\begin{abstract}

\etc{Leading approaches to fault-tolerant quantum computation dedicate a significant portion of the hardware to computational factories that churn out high-fidelity ancillas called magic states.  Consequently, efficient and realistic factory design is of paramount importance.   Here we present the most detailed resource assessment to date of magic state factories within a surface code quantum computer, along the way introducing a number of new techniques. We show that the block codes of Bravyi and Haah [\textit{Phys. Rev. A} \textbf{86}, 052329 (2012)] have been systematically undervalued; we track correlated errors both numerically and analytically, providing fidelity estimates without appeal to the union bound. We also introduce a subsystem code realisation of these protocols with constant time and low ancilla cost.  Additionally, we confirm that magic state factories have space-time costs that scale as a constant factor of surface code costs.  We find that the magic state factory required for post-classical factoring can be as small as 6.3 million data qubits, ignoring ancilla qubits, assuming $10^{-4}$ error gates, and the availability of long range interactions.}

\end{abstract}

\maketitle 

Architectures for quantum computers must tolerate experimental faults and imperfections, doing so in the most practical and efficient way.  One aspect of fault-tolerance is the use of error-correcting codes, which provides a storage method for protecting quantum information from noise.  To perform quantum computations, additional techniques are needed to ensure a universal set of quantum gates can be implemented fault-tolerantly.  Most error correcting codes natively allow fault-tolerant implementation of gates from the Clifford group, a non-universal set of gates.  Fully functional quantum computation is attained by adding the Toffoli or $\pi / 8$ phase gate to the Clifford group.  The prevailing proposal for performing these gates is to first prepare high-fidelity magic states, which are then used to inject a gate into the main computation.  These magic states are needed in vast quantities, and their preparation requires a significant portion of a device to operate as a dedicated magic state factory~\cite{BraKit05,Jones13,fowler13}.  Alternatives exist to the magic state paradigm~\cite{shor96,Knill96,Raussendorf06,bombin13b,Paetznick13,Anderson14,OConnor14}, but it is unclear whether they will be feasible substitutes due to worse thresholds~\cite{Brown16,bravyi15}.

\begin{table*}
\bgroup
\def\arraystretch{1.5}
\begin{tabular}{|p{2cm}|p{1.4cm}|p{1.4cm}|p{1.6cm}|p{1.6cm}|p{2cm}|p{2cm}|p{2cm}|p{2cm}|} \hline
\multicolumn{1}{ |c|}{\multirow{4}{*}{Problem} } 
& \multicolumn{2}{c|}{\multirow{2}{*}{Magic states required} } 
&  \multicolumn{2}{c|}{\multirow{2}{*}{\parbox{3.2cm}{spacetime overhead per magic state in qubit-rounds}}} 
&  \multicolumn{4}{c|}{\multirow{2}{*}{\parbox{8cm}{Physical qubits in factory (\textbf{and evaluation time}) required for time-optimal computation} }} \\

\multicolumn{1}{ |c|}{} & \multicolumn{2}{c|}{}  &  \multicolumn{2}{c|}{} & \multicolumn{4}{c|}{} \\ \cline{2-9}

\multicolumn{1}{ |c|}{} 
& \multicolumn{1}{ c| }{\multirow{2}{*}{Type}} 
& \multicolumn{1}{c|}{\multirow{2}{*}{Count}}  
& \multicolumn{1}{c|}{\multirow{2}{*}{$p_\mathrm{g}=10^{-3}$} }
& \multicolumn{1}{c|}{\multirow{2}{*}{$p_\mathrm{g}=10^{-4}$} }
& \multicolumn{2}{c|}{$p_\mathrm{g}=10^{-3}$, $t_{\mathrm{meas/ff}}=0.1t_{\mathrm{sc}}$}
& \multicolumn{2}{c|}{$p_\mathrm{g}=10^{-4}$, $t_{\mathrm{meas/ff}}=0.1t_{\mathrm{sc}}$} \\ \cline{6-9}
 
\multicolumn{1}{ |c|}{} 
& \multicolumn{1}{ c | }{} 
& \multicolumn{1}{ c | }{}  
&
&
&  $t_{\mathrm{sc}}=10^{-3}$ s
&  $t_{\mathrm{sc}}=10^{-5}$ s 
&  $t_{\mathrm{sc}}=10^{-3}$ s 
&  $t_{\mathrm{sc}}=10^{-5}$ s  \\ \hline \hline

\multirow{2}{*}{1000 bit Shor} 
& \multirow{2}{*}{ Toffoli }
& \multirow{2}{*}{$10^{10.60}$}
& \multirow{2}{*}{ $1.41\times10^{7}$ }
& \multirow{2}{*}{ $5.35\times10^{5}$ }
& \multirow{2}{*}{\parbox{2cm}{$1.73\times10^{8}$\textbf{   (6.6 weeks)}}}
& \multirow{2}{*}{\parbox{2cm}{$1.73\times10^8$  \textbf{   (11 hours)} }}
& \multirow{2}{*}{\parbox{2cm}{$6.30\times10^6$\textbf{   (6.6 weeks)}}}
& \multirow{2}{*}{\parbox{2cm}{$6.30\times10^6$\textbf{   (11 hours)}}}\\ 

&&&&&&&& \\ \hline

\multirow{2}{*}{2000 bit Shor }
&\multirow{2}{*}{ Toffoli }
&\multirow{2}{*}{$10^{11.51}$}
&\multirow{2}{*}{ $1.66\times10^{7}$ }
&\multirow{2}{*}{ $5.71\times10^{5}$ }
&\multirow{2}{*}{\parbox{2cm}{$2.18\times10^8$ \textbf{ (53 weeks)} } }
&\multirow{2}{*}{\parbox{2cm}{$2.18\times10^8$ \textbf{   (3.7 days)}}}
&\multirow{2}{*}{\parbox{2cm}{$6.97\times10^6$ \textbf{ (53 weeks)} } }
&\multirow{2}{*}{\parbox{2cm}{$6.97\times10^6$\textbf{   (3.7 days)}}} \\ 

&&&&&&&& \\ \hline

\multirow{2}{*}{4000 bit Shor }
&\multirow{2}{*}{ Toffoli }
&\multirow{2}{*}{$10^{12.41}$}
&\multirow{2}{*}{ $1.94\times10^{7}$ }
&\multirow{2}{*}{ $6.12\times10^{5}$ }
&\multirow{2}{*}{\parbox{1.9cm}{$2.50\times10^8$   \textbf{  (8 years)}}}
&\multirow{2}{*}{\parbox{2cm}{$2.50\times10^8$\textbf{   (4.2 weeks)}}}
&\multirow{2}{*}{\parbox{1.9cm}{$7.69\times10^6$   \textbf{  (8 years)}}}
&\multirow{2}{*}{\parbox{2cm}{$7.69\times10^6$ \textbf{  (4.2 weeks)}}}\\ 

&&&&&&&& \\ \hline

\end{tabular}
\egroup
\caption{\etc{The size and time requirements of some examples of magic state factories.  We consider an implementation of Shor's algorithm requiring $40 N^3$ Toffoli gates, which dominates the overhead.  We realise each of these gates using single Toffoli magic state or seven $T$ states in parallel~\cite{Selinger13}, whichever proves optimal.  In this algorithm, the Toffoli gates are all sequential, and so using time-optimal methods~\cite{Fowler2012} the fastest possible runtime is $40N^3 t_{\mathrm{meas/ff}}$  where $t_{\mathrm{meas/ff}}$ is the time taken to make a physical measurement and feed-forward the result to selectively perform a single qubit gate elsewhere in the quantum computer. The number of `physical qubits in factory' neglects qubit cost associated with measuring surface code stabilizers, and so for many architecture this number will be doubled. The variable} $t_{\mathrm{sc}}$ is the time taken to perform single round of the parallel stabilizer measurements of the surface code - a process involving four CNOT gates, two single qubit gates and a measurement. We assume throughout that $t_\mathrm{meas/ff}=0.1\cdot t_\mathrm{sc}$, which is reasonable for a distributed architecture such as ion traps.} 
\label{ShorTable} 
\end{table*}

Magic state factories use several rounds of distillation protocols, and several directions have been explored~\cite{Meier13,Bravyi12,Jones12,anwar12,Campbell12,campbell14} to improve efficiency over the original proposal which uses Reed-Muller codes~\cite{BraKit05}.  Notably, an $n \rightarrow k$ block protocol takes $n$ input magic states and output $k$ at higher fidelity, with higher ratios of $n$ to $k$ generally offering greater efficiency~\cite{Meier13,Bravyi12,Jones12}. These block protocols do require more complex circuits, but there has been limited investigation into the full resource cost of these protocols. One advance in this direction~\cite{fowler13} has shown that block protocols can be realized in constant time, independent of $k$, by braiding defects in the surface code.  Despite this, the same work found that efficiency improvements of block protocols were modest.  However, all previous work has taken a very pessimistic estimate of the fidelity of these protocols, leading to an overestimated cost.  We present several results improving resource costs and leading to a more optimistic outlook for realising magic state factories.

We present a method of realising the Bravyi-Haah block protocols~\cite{Bravyi12} in constant time without relying on braiding operations, providing a fast implementation to other architectures.  Braiding operations natively support control-NOT gates with multiple target qubits in constant time, which is the key step in the circuit reduction of Ref~\cite{fowler13}.  Our approach does not require availability of such powerful gates, but rather reduces time costs by borrowing concepts from subsystem codes~\cite{poulin05,bacon06,aliferis07,bombin10} and notions of gauge fixing~\cite{Bombin13,bombin13b,Anderson14}.  In subsystem codes, operations acting on many qubits are broken up into more pieces each involving fewer qubits, which we find enables greater parallelisation in realising Bravyi-Haah.  Our proposal is especially applicable to distributed architectures for quantum computation, and provides these devices with an approach to magic state distillation that is low in ancilla and time costs.

We also show that magic state factories can make more use of block protocols than previously thought by using new techniques to reduce and calculate the global output error. Consider a protocol outputting $K$ qubits with multi-qubit global error rate $\epsilon_g$.  All previous studies have used the union bound, also called Boole's inequality, to assert $\epsilon_g \leq K \epsilon$ where $K$ is the number of magic states and $\epsilon$ is the average error rate of single-qubit outputs, obtained by tracing the qubit from the multi-qubit state.  For an uncorrelated product state, $\rho_g = \rho^{\otimes K}$, we have $\epsilon_g = 1 - (1-\epsilon)^K$ and so to leading order $\epsilon_g = K\epsilon - O(\epsilon^2)$, so the union bound is a good estimate for small $\epsilon$.  However, the output of block protocols can be highly correlated. Boole's inequality holds for correlated states, but $\epsilon_g$ can be much smaller than $K \epsilon$.  In distillation protocols such as Bravyi-Haah, correlations are produced because block protocols reduce the probability of single errors, but pairs or clusters of errors can go undetected and lead to a correlated error pair.  If one error is present, it almost certainly has a partner.  As such, use of the union bound has lead to systematic over estimation of noise in Bravyi-Haah and other distillation protocols. Here we track these correlations through multiple rounds of a magic state factory.  We take as our starting point the triorthogonal codes for producing $\pi / 8$  magic states~\cite{Bravyi12} and a method of producing Toffoli magic states~\cite{eastin13,jones13b}.  Once we begin tracking correlations, it becomes apparent that quality control of the factory can be improved.  Detecting an error in one part of the factory can, due to correlations, indicate a likely undetected partner error elsewhere in the factory.  We introduce the notion of module checking whereby we discard magic states brought into disrepute by their correlated partners.   The enhanced fidelity of module checking is both analytically estimated and found by numerical Monte Carlo simulations, with excellent agreement between these methods.

\begin{figure*}[t]
    \includegraphics{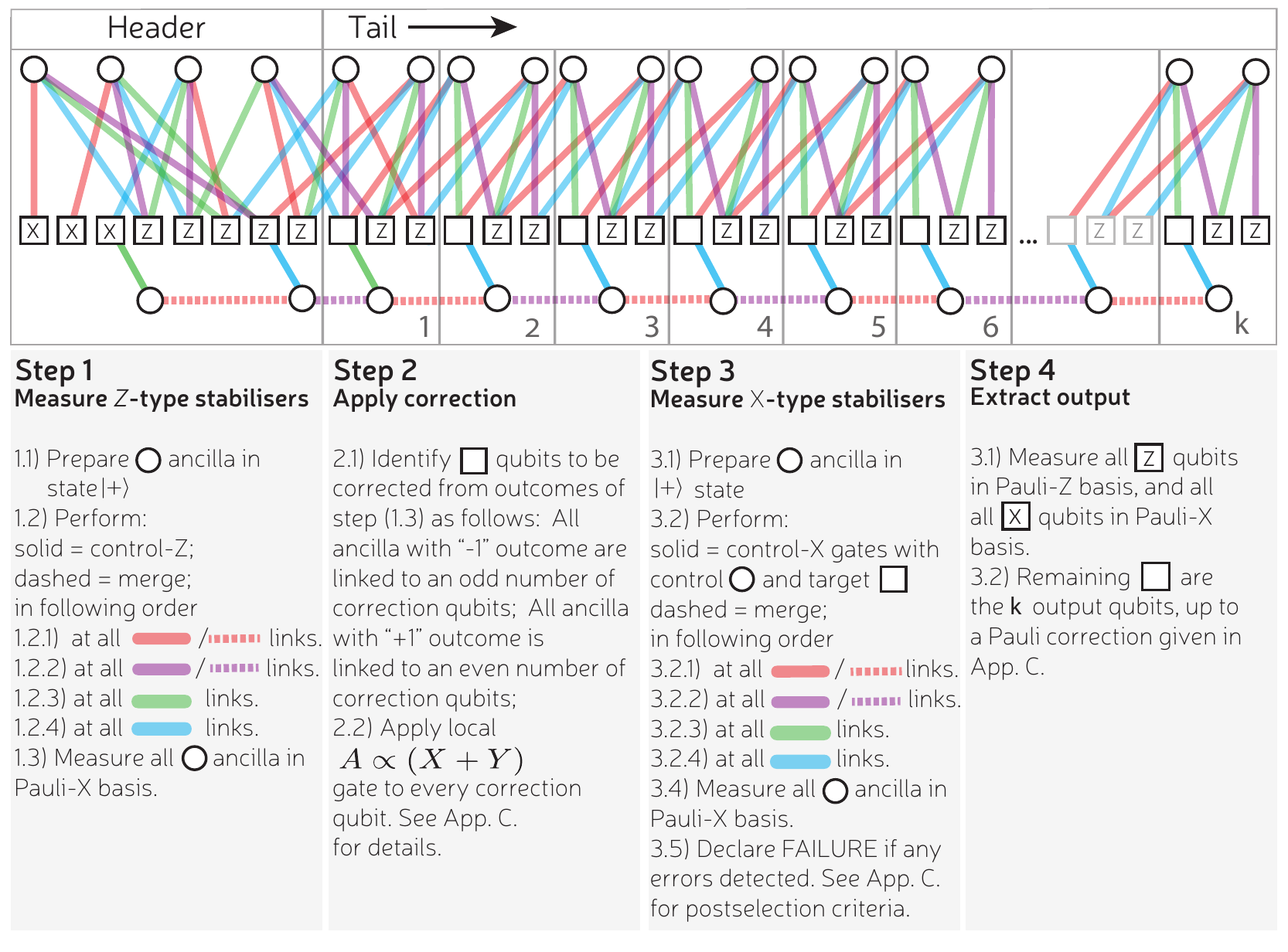}
    \caption{Explicit circuit for realising Bravyi-Haah $(3k+8) \rightarrow k$ block protocols for $k=2, 6, 10,14, \ldots$.  Squares indicate the $(3k+8)$ noisy $\ket{T}$ magic states to be distilled, and circles represent ancilla qubits used to effect measurements on the magic states.  Increasing $k$ does not increase the number of time steps in the protocol, but increases the number of qubits involved in a block.  As we increase $k$ we add qubits to the tail end, with the protocol translationally invariant along the tail. \textcolor{black}{We report that we have independently confirmed the validity of the protocol for $k=2$ by full wavefunction simulation, which further confirmed that all single errors are detected and all two errors processes lead to outputs with correlated errors.}}
    \label{BHprotocol}
\end{figure*}

\etc{The specification of the distillation protocol is just one aspect of designing magic state factories.   The size of magic state factories depends both on the distillation protocols and also the underlying error correction codes, and significant saving can be made by judiciously reducing the error correction code at earlier rounds of magic state distillation.  The balanced investment of qubits at each round of distillation uses small surface codes for low-fidelity magic states and larger surface codes for high-fidelity magic states. A small fraction of magic states in the factory will be high-fidelity and require much larger surface codes. Raussendorf \textit{et al.}~\cite{Raussendorf2007} argued that consequently magic state distillation can be achieved at a constant factor cost over surface code error correction, which we also observe and discuss.} 

\etc{Taking all factors into consideration, we present a blueprint for a factory capable of delivering enough magic states to solve large Shor's algorithm tasks, beyond the reach of classical computation, within a surface code quantum computer.} Our results are summarized in Table~\ref{ShorTable}, where we look at both the spacetime overhead required to produce a Toffoli magic state and the physical footprint of the factory required to produce magic states at an average rate that can just keep up with the `time optimal' surface code implementation of the algorithm~\cite{Fowler2012}, which we further discuss in Section~\ref{sec::overheads}.


\section{Realising block protocols}
\label{sec::realisation}

Many descriptions of distillation protocols are high-level, leaving open many aspects of how to implement these protocols. To assess the full resource cost, we require a low-level description of distillation protocols in terms of elementary operations, such as one and two qubit gates, preparations and measurements.  We call such a description a realisation of a protocol, and a given protocol can have different realisations with varying costs.  

This section presents a realisation of the Bravyi-Haah protocols, giving explicit instructions presented in Fig.~\ref{BHprotocol}.  The protocol is presented as a four step process acting on a collection of $n$ noisy $\ket{T}$-states, where $\ket{T} \propto \ket{0}+\exp(i \pi/4) \ket{1}$.  The first step uses ancillas to measure operators composed of Pauli-$Z$ operators, and the second step applies a correction dependent on the measurement outcomes.  The third step uses ancillas to measure operators composed of Pauli-$X$ operators, with only certain measurement outcomes kept.  After a successful third step, the $k$ output magic states are within an encoded state and delocalized across $3k+8$ sites. The fourth step uses measurements to localise the output qubits to specific sites.  All these steps are detailed in the figure, and show how the multi-qubit measurements are broken down into ancilla preparation, two-qubit gates, and single qubit measurements.  Observe in Fig.~\ref{BHprotocol} that each multi-qubit measurement involves only four entangling gates, with each such gate designated a distinct colored link.   Therefore, the measurement is a four qubit operator.  This is called the weight of the operator.  When using a single ancilla to measure a stabilizer of weight $m$, we need $m$ time steps to perform the required controlled-gate operations.  The low, and constant, weight of our measurements gives the realisation a constant time cost.  More commonly, the Bravyi-Haah  protocol is presented as requiring only 2 measurements of $X$-type observables that have weight $2k+4$,  implying a potentially expensive time cost. However, the concept of gauge subsystem codes provides a method whereby such complex measurements can be broken down into a larger number of simpler measurements~\cite{poulin05,bacon06,aliferis07,bombin10}.  In App.~\ref{SubSystem} we present a rigorous demonstration that the Bravyi-Haah code can be viewed as a subsystem code, and using a combination of gauge fixing and a cat state ancilla we create a circuit of depth 4 for both X and Z measurement sets. We call this general approach gauge-MSD\etc{, where MSD is short for magic distillation}. We remark that constant time realisation was also found Fowler \textit{et. al.}~\cite{fowler13}, but was tied to a monolithic braiding architecture.
  
The time complexity of our realisation is simply
\begin{equation}
 t_{\mathrm{block}} = 8 t_{\mathrm{cnot}} + t_{\mathrm{A}} + 2 t_{\mathrm{prep}} + 3 t_{\mathrm{measure}} \sim 8 d t_{\mathrm{sc}},
\end{equation}
where $t_{\mathrm{cnot}}$ is the 2-qubit gate (e.g. CNOT) time, $t_{\mathrm{A}}$ \textcolor{black}{is the gate} time for single qubit $A=(X+Y)/\sqrt{2}$ rotation, $t_{\mathrm{prep}}$ is the single qubit preparation time, and $t_{\mathrm{measure}}$ is the single qubit measurement time.  Here, all operations are applied fault-tolerantly to logical qubits within an error correcting code.  Therefore, the time scales are for fault-tolerant gates. We will assume throughout that logical CNOT gates are applied transversally and thus take time $t_\mathrm{cnot}=t_\mathrm{g}+d \times t_\mathrm{sc}$ where $t_g$ is the time for the physical gate, $d$ is the code distance and $t_\mathrm{sc}$ is the time for a round of stabilizer measurements. \etc{Therefore, for large distances the time is dominated by $8 d t_{\mathrm{sc}}$}. \textcolor{black}{We will also take this as the time cost of a merge operation~\cite{horsman12,landahl14} which we use to project two ancillary qubits into the even or odd parity subspace.  We assume entangling gates can be performed in parallel, but a qubit can only participate in one gate at a time. We allow entangling gates to be long-range as is feasible within distributed architectures for quantum} computing~\cite{BK01a,JTSL02a,ODH01a,Moehring07,Camp08,Camp2010review,Hanson13,Nickerson14}, though this may be relaxed at only a modest increase in resources.  

 Our realisation uses a number of ancilla qubits, so that in addition to the $n=8+3k$ qubits being distilled we also \textcolor{black}{use $n_{\mathrm{anc}}=3k+6$ ancilla qubits, giving $n_{\mathrm{tot}}=6k+14$ logical qubits in total.   These ancillas appear as circles in Fig.~\ref{BHprotocol}.  With these logical qubits  encoded in a distance $d$ toric code, the total qubit cost is $N_{\mathrm{tot}}=(6k+14)d^2$.  Therefore, the total space-time cost amounts to $N_{\mathrm{tot}}t_{\mathrm{block}} \sim (48k+112)d^3$, which compares well against other approaches (see App.~\ref{App::compare}).}

When using the toric code $A$ gates are also not naturally fault-tolerant and thus require their own process of state distillation and injection. The time $t_A$ to implement such a gate is therefore the time taken by the logical CNOT which teleports the gate into the computation plus any Clifford correction that is required, although this can be rolled into a later Clifford operation. The resources required of the state distillation of the $A$ are far less than that of the $T$-gate, and as the ancilla resource is reusable for many $A$ gates~\cite{Jones12,Anderson12}. As such we neglect the overhead of these gates as a small additional overhead to the main process of $\ket{T}$ distillation.



\section{Overview}

\subsection{Blocks, branches and modules}

We begin by introducing some helpful vocabulary for describing magic states factories. Efficient distillation uses $n \rightarrow k$ block protocols, that take $n$ noisy $\ket{T}\propto \ket{0}+\exp(i \pi/4) \ket{1}$ and with some probability output $k$ states of a higher fidelity.  Such a process we call a block, and the previous section described the details of the inner working of such a block.  Now we treat each block as a black box with known relations between input and output, and consider how these blocks are composed together. 

\begin{figure}[t]
\includegraphics{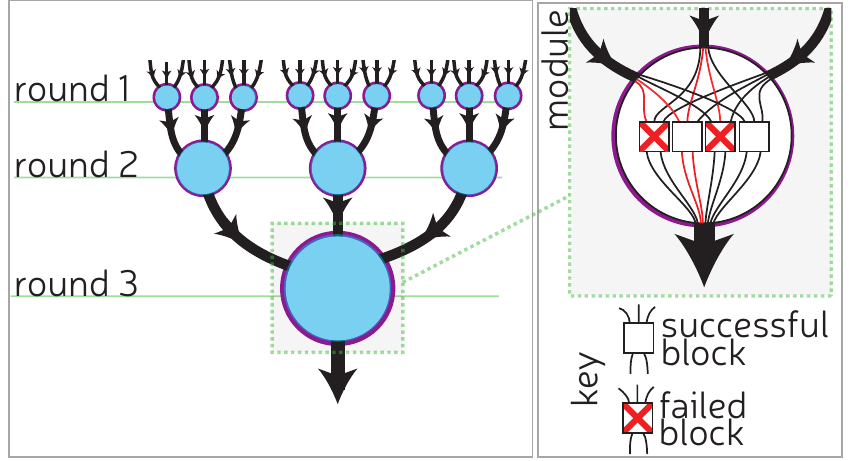}
\caption{The a tree structure of many rounds of distillation, with branches (directed black lines) that merge at branching points that we call modules. The thickness of the branch increases with each round.    The figure shows a fictitious scheme where $n_l=3$ and $k_l=2$ for all rounds.  (inset) The structure within a module. Incoming branches contain many qubits, here this is shown to be 4. These qubits undergo a permutation $\sigma$ and are feed into an instance of a block of a distillation protocol shown as a square. Here the 3 incoming branches carry 4 qubits and so we need 4 instances of a $3 \rightarrow 2$ protocol. We use a fictitious protocol to keep the numbers low enough to illustrate clearly. Each of the 4 blocks output 2 qubits and these are merged into a branch of 8 qubits feed into a later module. A pernicious error pattern is shown in red, which is detected in 2 of distillation blocks, marked with crosses, but goes undetected in a third block.}
\label{FIGtree}
\end{figure}

Distillation protocols have many levels forming a tree-like structure with many branches that merge at points we call modules, shown in Fig.~\ref{FIGtree}. Branches contain many qubits, which are potentially correlated.  However, the inputs to block protocols must not be correlated, so each qubit in a branch must be fed into a different block.  Therefore, as we enter a module, a branch of $B_l$ qubits is split up so that each qubit enters a different block.  If each block implements a $n_l \rightarrow k_l$ protocol, then the whole module can be thought of taking $B_l n_l$ inputs to $B_l k_l$ outputs.  This entails that $B_{l+1}=B_lk_l$ and that each module has $n_l$ branches feeding into it.  Initially, branches are single qubits, $B_1=1$, and so $B_l = \prod_{1 \leq j < l} k_j$.  This module-branch structure is common to all proposals to date.  Such explicit terminology has not previously been introduced but rather been left as an implicit consequence of statements about correlation avoidance.  Establishing clear vocabulary about this structure is important as we delve into the effect of postselecting at different levels on this structure.  Previous protocols have considered whether individual blocks succeed or fail, we call this \textit{block checking}.  Below, we outline why it can be advantageous to postselect on the level of the modules, which are collections of blocks. We propose an additional quality check, so that the whole module is discarded whenever any of its blocks fail. We call this \textit{module checking}.

A block will always detect a single incoming error, but might fail to detect a pair of errors.  When a block detects an error, it indicates the presence of damaged branches, and since errors cluster together within branches, this increases the likelihood of errors in other blocks throughout the module.  Consider when two branches fail each with a correlated error pair, the first branch sends damaged qubits to blocks 1 and 2, and the second branch sends damaged qubits to blocks 2 and 3.   Since blocks 1 and 3 each received a single erroneous qubit, they will detect them.  But block 2 receives a pair of errors, so they may go undetected.  A simplified illustration of this process is shown in Fig.~\ref{FIGtree}.  Module checks improve fidelity by preventing these processes from degrading the output fidelity.  Even with module checks, it is possible for a pair of corrupted branches to go undetected, but both branches must carry exactly the same pattern of errors, which is a very rare occurrence. 

\section{The G-matrix formalism}

Bravyi and Haah introduced a matrix description of their $n \rightarrow k$ block protocols for $\ket{T_0}$ state distillation. The so-called $G$-matrix is split into two submatrices $G_1$ and $G_0$, with $G_0$ describing the postselection criteria and $G_1$ \textcolor{black}{accounting} for how input qubits are related to output qubits.  For distillation of $\ket{T_0}$ magic states, the $G$ matrix must have the property of triorthogonality that the reader can learn about in Ref~\cite{Bravyi12}. Rather here we give a pragmatic account of the the block's performance.  We use $\ket{T_0} \propto \ket{0}+e^{i \pi/4} \ket{1}$ for a magic state and $\ket{T_1}=Z \ket{T_0}$ for the orthogonal state with a $Z$ error.  A pure multi-qubit state $\ket{T_{x_1}} \ket{T_{x_2}} \ldots \ket{T_{x_n}}$, we concisely represent with the vector $x=\{ x_1, x_2, \ldots x_n \}$.  If we apply a block protocol to state $x$, the block succeeds (detecting no errors) if $G_0 x = 0 \pmod{2}$ where the maths is performed modulo 2.  When successful, the block outputs a state $y=G_1 x$.  Noisy magic states will be some probabilistic ensemble over $x$, with probability $\Pr(x)$.  \etc{The protocol will detect no errors and output state $\ket{T_{y_1}} \ket{T_{y_2}} \ldots \ket{T_{y_k}}$ with (unnormalized) probability
\begin{equation}
\label{PrY}
    \Pr_{\mathrm{unnorm}}( y ) =  \sum_{ \{ x :  G_0 x = 0 , G_1 x = y   \} } \Pr( x ) .  
\end{equation}
The total success probability is captured by the sum over all possible output states, so
\begin{align}
\label{Psuc} \nonumber 
    P_\mathrm{\mathrm{suc}} & = \sum_{y}  \Pr_{\mathrm{unnorm}}( y ) \\  \nonumber \nonumber
 & = \sum_{y}  \sum_{ \{ x :  G_0 x = 0  , G_1 x = y \} } \Pr( x ) \\
 & =   \sum_{ \{ x :  G_0 x = 0  \} } \Pr( x ) .   
\end{align}
Conditioned on success, the normalized distribution on output states is $\Pr_{\mathrm{out}}( y ) = \Pr_{\mathrm{unnorm}}( y )  /  P_\mathrm{\mathrm{suc}} $.}  Given an explicit form for $G_0$ and $G_1$, this completes the black box picture of block protocol performance.  These formulae form the basis upon which we build both our analytic and numerical analysis in following sections.

The $G$-matrix formalism of Bravyi and Haah has been significantly generalized~\cite{Camp16b,camp16d}.  This extension provides protocols that convert noisy $T$ magic states into another species capable of injecting complex multi-qubit circuits.  Included in this framework are protocols, based on $G$-matrices, which provide resources for implementing Toffoli gates.  Protocols independently proposed by Jones~\cite{jones13b} and Eastin~\cite{eastin13} realized error suppressed Toffoli gates, and here we consider a variant based on $G$-matrices that we discuss further in App.~\ref{Toff}.  All three variants perform identically when we use block checking.  However, with the $G$-matrix formalism we can again use module checking to track correlations and achieve superior error suppression.  This is just one additional application of module checking, the technique can be deployed in conjunction with the general class of protocols introduced in Ref.~\cite{Camp16b,camp16d}

\section{Analysis of module checking}
\label{sec::module}

We present a method of tracking the leading order errors, accounting for correlations, through many rounds of module checked protocols.  At each level of distillation the protocol is characterized by a function $\ff$ that summarises how well it tolerates leading order errors 
\begin{defin}
\label{distillationFunction}
For every distance 2 $G$-matrix code that \textcolor{black}{distills} $n \rightarrow k$ qubits, we define a function $\ff : \mathbb{Z}_2^k \rightarrow \mathbb{Z}$ taking values
\begin{equation}
    \ff (y) := \# \{ y : |x|=2, G_0 x = 0 , y=G_1 x \}  ,
\end{equation}
where $| \ldots |$ is the weight $|y|=\sum_j y_j$, and $\#$ counts the number of elements in a set $\{ \ldots \}$.  In other words, the value $\ff (y)$ counts the number of inputs $x$ such that:
\begin{enumerate}
    \item they are weight 2 (formally $(|x|=2)$) ; and
    \item they are undetected by the protocol (formally $G_0 x = 0$); and
    \item give $y$ as output (formally $G_1 x = y$).
\end{enumerate}
\end{defin}

\begin{table*}
\begin{tabular}{|c|c|c|c|c|} \hline
 level 1 & level 2  & $C_l$ &  $\lim_{k \rightarrow \infty} C_l$ &$\lim_{k \rightarrow \infty} k_1k_2\epsilon_{\mathrm{BH}}/\epsilon^4$ \\ \hline \hline
 $ \mathrm{BH}_{k_1}$ & $ \mathrm{BH}_{k_2}$ & $\left( 16 + \frac{9}{2}k_1(k_1 - 1) \right)\left( 4 + \frac{3}{2}k_2(k_2 - 1) \right) $ & $\frac{27}{4} k_1^2 k_2^2$ &$27k_1^3k_2^2$  \\
  $ \mathrm{Tof}$ & $ \mathrm{BH}_{k}$ & $ 112 \left( 4 + \frac{3}{2}k(k - 1) \right) $ & $168 \cdot k^2$ & $2352\cdot k^2$\\
  $ \mathrm{BH}_{k}$ & $\mathrm{Tof}$ & $  \left( 16 + \frac{9}{2}k(k - 1) \right) 28 $ & $126 \cdot k^2$ & $252\cdot k^3$\\ \hline
 \end{tabular}

 \begin{tabular}{|c|c|c|c|c|c|} \hline
 level 1 & level 2  & level 3  & $C_l$ &  $\lim_{k \rightarrow \infty} C_l$ &$\lim_{k \rightarrow \infty} k_1k_2k_3\epsilon_{\mathrm{BH}}/\epsilon^8$  \\ \hline \hline
 $ \mathrm{BH}_{k_1}$ & $ \mathrm{BH}_{k_2}$ & $ \mathrm{BH}_{k_3}$ & $\left( 256 + \frac{81}{2}k_1(k_1 - 1) \right) \left( 16 + \frac{9}{2}k_2(k_2 - 1) \right) \left( 4 + \frac{3}{2}k_3(k_3 - 1) \right) $ & $ 273.375 \cdot k_1^2 k_2^2 k_3^2$ &$2187\cdot k_1^5k_2^3k_3^2$\\
  $ \mathrm{Tof}$ & $ \mathrm{BH}_{k_1}$ & $ \mathrm{BH}_{k_2}$ & $ 1792  \left( 16 + \frac{9}{2}k_1(k_1 - 1) \right) \left( 4 + \frac{3}{2}k_2(k_2 - 1) \right) $ & $ 12096 \cdot k_1^2 k_2^2$&$28^4\cdot 3^3\cdot k_1^3k_2^2$ \\
   $ \mathrm{BH}_{k_1}$ & $ \mathrm{BH}_{k_2}$ & $\mathrm{Tof}$ & $\left( 256 + \frac{81}{2}k_1(k_1 - 1) \right) \left( 16 + \frac{9}{2}k_2(k_2 - 1) \right) 28 $ & $ 5013 \cdot k_1^2 k_2^2$ &$20412\cdot k_1^5k_2^3$ \\ \hline
 \end{tabular}
 \caption{The leading coefficient $C_l$ for a variety of protocols with 2 and 3 levels of distillation.  For clarity we also show $C_l$ in the large block limit $(k \rightarrow \infty )$.  When we write $\mathrm{BH}_{k}$, we implicitly assume $k>2$, as the results differ slightly for the $k=2$ case. The final column shows the ratio between the union bound estimate made by utilising the reduced error rate on a single qubit $\epsilon_{\mathrm{BH}}$ made by Bravyi and Haah and the corresponding estimate of the global error rate given by $C_l\epsilon^{2^{l}}$. It can be seen that the benefit (in error rate) of module checking scales with both $k$ and the number of rounds of distillation.}  
\end{table*}

Since $\ff(y)$ counts the number of lowest-weight errors leading output $y$, the total error rate for 1 round of distillation can be simply estimated as
\begin{equation}
    \epsilon_{g} = \left( \sum_y \eta(y) \right) \epsilon^2 + O(\epsilon^4),
\end{equation}
Counting errors over many rounds is a more subtle problem, but we find that $\ff$ still provides sufficient information to perform this calculation.  If each round can even use a different protocol, we label the corresponding function with a subscript. We now state our key result
\begin{theorem}
\label{DistFuncThm}
Consider $L$ rounds of distillation with module checking, with associated functions $\ff_1, \ff_2, \ldots \ff_L$.   Such a protocol outputs a multi qubit magic state where the $l^{\mathrm{th}}$ level modules succeed with probability
\begin{equation}
    P_{\mathrm{\mathrm{suc}}, l} \simeq \frac{A_l + B_l}{(A_{l-1}+B_{l-1})^{n_{l}}} 
    \end{equation}
and output states have global infidelity
\begin{equation}
    \epsilon_g^{(l)} \simeq \frac{B_l}{A_l+B_l} .
    \label{eqn::eglo}
\end{equation}
where we have made use of the following
\begin{align}
\label{ABCexpressions}
    A_l & = (1-\epsilon)^{(n_1 n_2 \ldots n_l)} , \\
    B_l & = C_l \epsilon^{2^l}(1-\epsilon)^{(n_1 n_2 \ldots n_l - 2^l)} , \\
    C_l & = \prod_{j=1}^{l} \left( \sum_v \eta_{j}(v)^{2^{l-j}} \right).
\end{align}
\end{theorem}
The most important quantity in the theorem is $C_l$.  After two rounds $C_2$ is simply $(\sum_y \eta_1(y)^2)\cdot(\sum_y \eta_2(y))$.  After $l$ rounds, $C_l$ is a product of $l$ terms.  One may approximate the theorem to leading order $\epsilon_g^{(l)} \sim C_l \epsilon^{2^l}$.  However, our later numerical investigations found that such an approximation was too coarse, and we really need the slightly more complex form given in the theorem.  We postpone proof of this result to App.~\ref{proof}.

For the Bravyi-Haah protocols it is easy to verify the following
\begin{equation}
    \ff_{\mathrm{BH}_k}(y) = \begin{cases}
     3 &, |y|=2; \\
     4 &, |y|=k; \\
     0 &, \mathrm{otherwise}.
 \end{cases}
\end{equation}
so that for $k>2$ we have
\begin{equation}
    \sum_y \ff_{\mathrm{BH}_k}(y)^{m} = 4^m + 3^m \binom{k}{2} = 4^m + 3^m \frac{ k (k-1)}{2} .
\end{equation}
where the binomial factor arises in counting the number of $y$ where $|y|=2$.  When $k=2$, we have a special case as then the $|y|=2$ terms and $|y|=k$ terms are the all same, and so
\begin{equation}
    \sum_y \ff_{\mathrm{BH}_2}(y)^{m} = 7^m.
\end{equation}
 For the Toffoli protocol we have
\begin{equation}
    \ff_{\mathrm{Tof}}(y) = 4 ,  y \neq 0.
\end{equation}
and so 
\begin{equation}
  \sum_y   \ff_{\mathrm{Tof}}(y)^m = 7 \cdot 4^m .
\end{equation}
Given expressions for $\ff(y)^m$ we can compose these protocols anyway we wish and obtain an estimate of the global error rate as given in Thm.~\ref{DistFuncThm}.  
  

We perform numerical Monte Carlo simulations by sampling from the probability distribution of the raw magic states, where $\Pr(x)=\epsilon^{|x|}(1-\epsilon)^{N-|x|}$, and track its evolution through the magic state factory. To our knowledge, this is the first such numerical investigation of correlations inside a magic state factory.

Our simulations track the progress of potentially damaging input error configurations of the initial raw magic states though the factory. Direct Monte Carlo simulation is possible for 1-2 rounds of distillation, but with 3 or more rounds the failure events become so rare that brute force simulation fails to provide statistically meaningful data.  Rather we use a  novel ``rare events" technique that preselects cases with two or more corrupt branches. A full description of the method employed can be found in Appendix~\ref{simulations}. We thus investigate the performance of a factory which makes use of module checking for a range of raw magic state error rates between 0.1\% and 1\%. We find that the leading order analytic estimates match well with the numeric results, with the difference between the two being of the order of a few percent in the investigated parameter regime. In fact wherever the block protocols are involved, if $k\leq14$ we find the percentage difference in between the numerical simulation and analytic estimate is $<10\%$. This discrepancy between the analytic estimate and numerical simulations is not visible on a log-log plots presented in App.~\ref{simulations}.

The cost of a protocol is the average number of raw magic states consumed to produce one higher fidelity magic state. For an $n\rightarrow k$ protocol, which takes in states with error $p$, this is 
\begin{equation}
\mathcal{C}(p)=\frac{n}{kP_s(p)}
\end{equation}
For $l$ rounds of distillation we have $\mathcal{C}_l(p)=\prod_{i=1}^{l-1}\mathcal{C}_{i}(p_i)$.

\begin{figure*}
\centering
\includegraphics{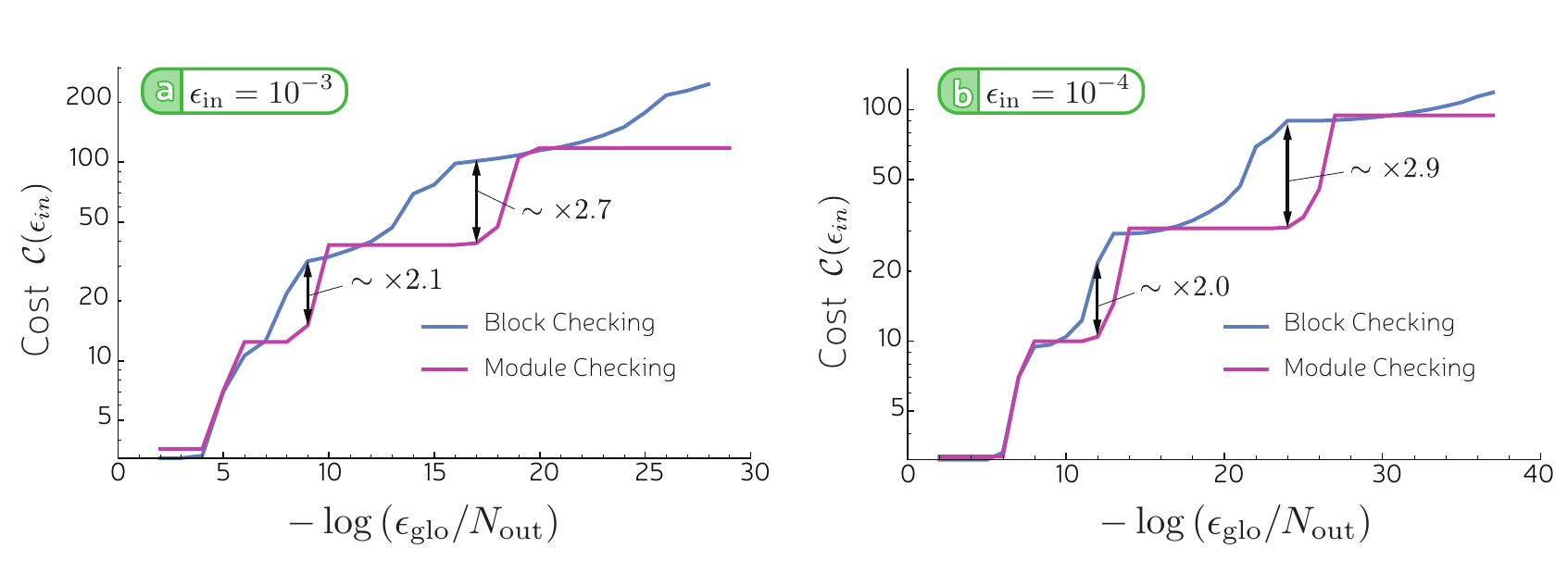}
\caption{(a) \& (b) The cost $\mathcal{C}$ of the Bravyi-Haah protocols utilising both block checking and module checking. It can be seen that only around the transitions to an additional level of distillation is block checking very slightly preferable.}
\label{YieldFig}
\end{figure*}

Fig.~\ref{YieldFig} compares the cost of a Bravyi-Haah magic state factory, with block checking and module checking, showing the minimum cost achievable for given output error. For output error rate and success probability we use known expressions for block checking, and for module checking the success probability and global error given by Thm.~\ref{DistFuncThm} with an estimate on the reduced error rate on a single qubit given by the global error rate estimate divided by the total number of qubits in this factory's output.

We find that module checking is superior to block checking for a large proportion of target error rates, and can use up to four times fewer raw magic states in some regimes. However,  near a transition from $j$ to $j+1$ rounds of distillation,  module checking loses it advantage and may even be slightly outperformed. The best error rates that can be achieved for a given number of rounds use low $k$ block codes, for these the benefit in global error rate of module checking over block checking is smaller (see Table II) while the success probability is of course much inferior. Above the transition higher $k$ values are being used, for which the success probability is much lower, and this washes out the benefit of the superior error suppression over block-checking, which has a higher success probability.  In these regimes, as one might expect, module checking is not the optimal approach. In those regions between the transitions, module checking allows use of higher $k$ protocols, which are more efficient,  to achieve the error rates of lower $k$ block checked protocols.

\section{Factory overhead analysis}
\label{sec::overheads}

While the `cost' is a useful guide to the performance of a distillation protocol, it fails to capture several important features of real magic state factories. The very purpose of magic state distillation is to supplement the shortcomings of a particular error correcting code, which is to say that a magic state factory is implemented at the level of logical qubits, with the quantum information already encoded. As such, the more relevant question is not how many noisy input magic states are required per output state, but rather the number of the physical qubits that will build up such a factory and the rate at which the distilled magic states are produced. Both of these numbers are of key importance to determining the size of a quantum computer and the run time of an algorithm. A single number, the `spacetime overhead' of the magic state factory captures these both as a figure of merit, which was also studied in Ref~\cite{fowler13}. In this section we present a comparison based on the spacetime overhead of implementing a magic state factory in a surface code quantum computer, utilising module checking, where appropriate, and our novel implementations of the Reed-Muller, Bravyi-Haah and Toffoli protocols.

We consider a number of issues in this estimate of the footprint of a magic state factory, including
\begin{enumerate}
 \item 	\textit{balanced investment}: the use of smaller surface codes during early rounds of magic state distillation;
 \item 	\textit{clock rate zoning}: cycling through distillation attempts faster during early rounds of magic state distillation;
\end{enumerate}

We will assume throughout that we use the method outlines by Li in Ref~\cite{Li2015} to inject the initial raw magic state into a logical surface code. We will therefore assume throughout that the initial error rate on a magic state before distillation is $\epsilon_\mathrm{in} = 0.4 p_\mathrm{g}$. We also use the `rotated lattice' surface codes~\cite{horsman12}, such that a distance $d$ surface requires $d^2$ physical data qubits. Of course a practical surface code requires the use of physical ancilla qubits to make the stabilizer measurements of the code, we leave this as an extra multiplicative factor to be applied to our overhead calculated here, as different physical realisations have different requirements. We estimate the surface code distance required to protect a logical qubit \textcolor{black}{up to} error rate $P_l$ using the relation $P_l(d,p_\mathrm{g})=d(100p_\mathrm{g})^{\frac{d+1}{2}}$~\cite{fowler13}.

Given an algorithm, and some implementation of it, the number of magic states required is determined in advance. For a given distillation protocol we must then determine whether its magic state factory is capable of producing magic states of fidelity great enough that there is a high probability that none of the magic states required for the algorithm fails. Thus we set a target global error rate that our factory must achieve.

\begin{equation}
\epsilon_\mathrm{target}= 1-(P_\mathrm{suc,alg})^{1/N}
\end{equation}
where $P_\mathrm{suc,alg}$ is desired success probability of our algorithm (i.e. the probability that every non-Clifford gate works) and $N$ is the number of successful iterations of the factory needed to produce the desired number of magic states. For example, a magic state factory which utilises 3 rounds of Bravyi-Haah distillation with a $k$ value of $10$ in each round will produce $10^{15}$ $\ket{T}$ states after $10^{12}$ successful iterations. A 90\% success probability for the algorithm as a whole then implies $\epsilon_\mathrm{target}= 1.05\times10^{-13}$.
We must then check that the factory is capable of producing an output of this quality. If the factory is module checked then this `10-10-10' factory has a global error rate $\epsilon_\mathrm{glo}=2.3\times 10^{-16}$, making this a valid protocol. However the estimate of the global error rate of a block checked factory gives $10^3 \times \epsilon_{BC}\sim10^{-11}$ so this would not be a valid factory for this task.

\subsection{Balanced investment}
 
Once we have established that a factory is valid we can calculate the spacetime overhead per magic state that it produces. This is done by: calculating the distance of surface code required at each level of distillation $d_i$; the length of time in surface code cycles that each round of distillation will take $T_i$ and the number of logical qubits $Q_i$, including logical ancillas, required at each level of the factory.

\begin{figure}
	\includegraphics{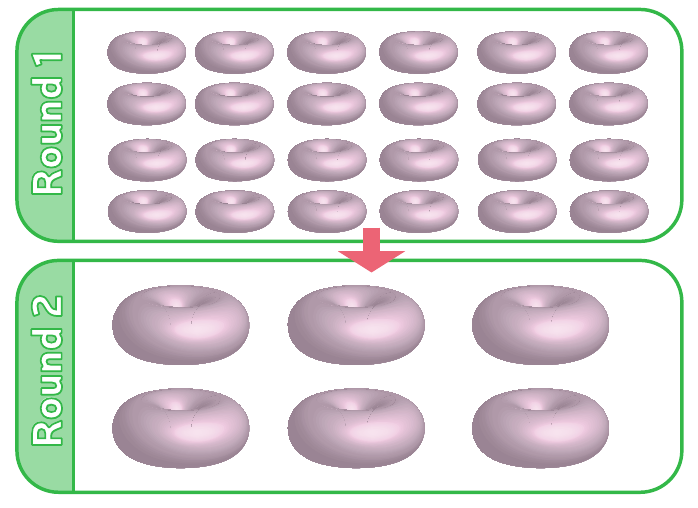}	
	\caption{The concept of balanced investment. During the initial rounds of distillation smaller surface codes can be used to encode the logical information. The factory requires only logical surfaces of the distance corresponding to the target error rate of the round in which that qubit is involved. This target in each round will depend on the final error target, the protocol being used to achieve it and the error rate of the raw state.}
\end{figure}

Determining the distance required at each level requires slightly different methods depending on whether module or block checking is used. To obtain the benefits of module checking we cannot make full use of balanced investment at the lower levels, as to do so would inject noise at a rate comparable or greater than the rate of correlated error. We can estimate the total global error output in the output of a $n\rightarrow k$ protocol as $\epsilon_\mathrm{tot} \sim \epsilon_\mathrm{glo} + k\epsilon_\mathrm{enc}$, where $\epsilon_\mathrm{enc}$ is the random error rate resulting from size of the logical encoding chosen. As such we determine the distance of the code at the intermediate levels based on a desired error rate $\epsilon_{\mathrm{enc}}$ to be $0.1\cdot\epsilon_\mathrm{glo}/k$ to ensure that the error due to each qubits finite encoding is much less than the reduced error $\sim \epsilon_\mathrm{glo}/k$ on that qubit. The final output of the factory can then encoded according to the $\epsilon_{\mathrm{target}}$.

\begin{figure*}[t]
\centering
\includegraphics[width=2.05\columnwidth]{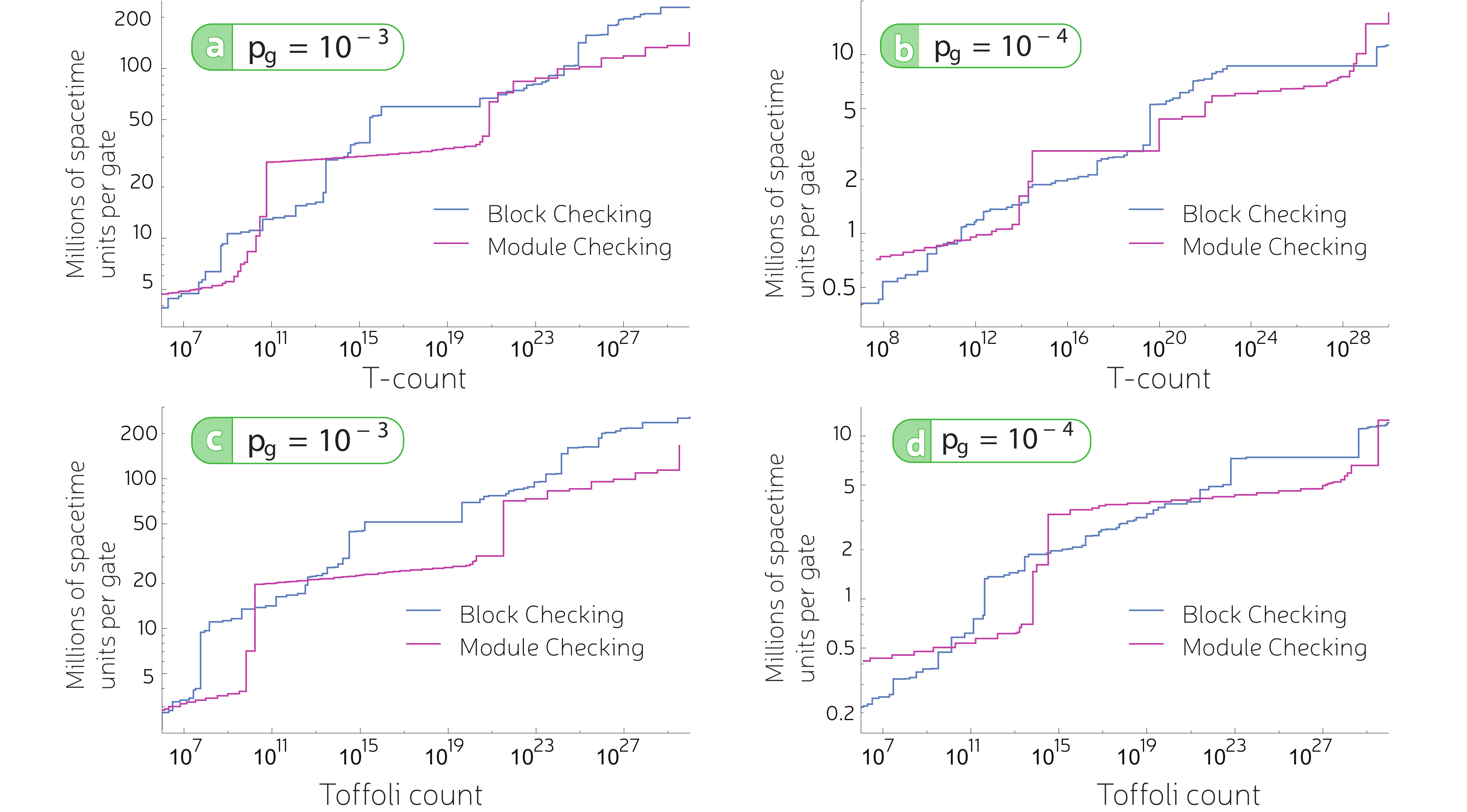} 
\caption{Spacetime overheads of producing both (a) \& (b) $\ket{T}$ states and (c) \& (d) $\ket{\mathrm{Toff}}$ states. Note that module checking is more beneficial when one of the rounds of distillation is Toffoli.}
\label{spacetimeoverheads}
\end{figure*}

For a block checked factory (or the original 15-to-1 Reed-Muller protocol) we do not have worry about `protecting' correlated errors. This means we can work backwards from our error target to determine an efficient balanced investment of qubits. For a local target error of $p_\mathrm{top}= 10^{-14}$ the top level of distillation needs $ v P_L(d,\epsilon_\mathrm{enc}) < 0.1\times 10^{-14}$ where $v$ is the spacetime volume of a single block of distillation; which is the number of surface code qubits in the block multiplied by the number of times they undergo $d$ rounds of surface code stabilizer measurements. Again we use a distance corresponding to a lower error rate by a factor of 10 to suppress any error injected by the logical circuitry above that which is left over after distillation. The distance required of the next level down can then be determined by, in this example $p_{i-1} =\sqrt[3]{p_\mathrm{top}/35}$ for the 15-to-1 protocol, which gives distance $d_{i-1}$. And so on until $p_{i}>p_\mathrm{in}$.

The length of time $T_i$ for each round of distillation can be simply determined by the protocol used and how many times we attempt it before abandoning the round. We assume that measurements can be completed in one time step, and the time scale of the CNOTs, preparation and A gates is dominated by the requirement for $d$ rounds of stabilisation afterwards. Therefore we let the time for each of these be $d\times t_{sc}$. The time taken to implement the distillation protocol in round $i$ is then,

\begin{equation}
    \tau_i = \begin{cases}
     11t_{\mathrm{sc}}\times d,& \text{round } i \text{ uses Bravyi-Haah} ; \\
     12t_{\mathrm{sc}}\times d,& \text{round } i \text{ uses Toffoli}. \\
     13t_{\mathrm{sc}}\times d,& \text{round } i \text{ uses Reed-Muller} .
 \end{cases}
\end{equation}

\subsection{Clock-rate zoning}
Balanced investment also allows another advantage. In the context of surface code computing, the distance of the code is not only relevant for the spatial dimensions of the computer, but also the execution time. A surface code of distance $d$ must undergo $d$ rounds of parallel stabilizers measurements to protect from measurement errors. As such, the time taken for a logical operation is proportional to $d$ (except those which can be handled in software) and therefore using balanced investment the initial rounds of distillation will take less time. In the hypothetical case that a round of distillation leads to a squaring of the input error rate, this would correspond to a doubling of the code distance required by the next round (by the exponential suppression of error with distance of a sub threshold surface code). Therefore one can repeat the first round of distillation twice in the time taken for the second round of distillation to be completed. This increases the chance that all the necessary magic states for the second round will have been produced in time for the next round of distillation, without the need to decrease the rate of the factory. As such the time taken for a round to complete in the distillation factory is $T_i=\tau_i t_i$ where $t_i$ is the number of attempts that you allow at each round. In all our simulations, any 'idle time' that the qubits experience is counted towards the spacetime cost.

\subsection{Numerical simulations}

We have therefore arrived at the expression for the full spacetime volume $\mathcal{V}$ in qubits-rounds occupied by a factory

\begin{equation}
\begin{split}
\mathcal{V}(\epsilon_\mathrm{in},p_\mathrm{g},\mathcal{N},\{k_i\},\{t_i\},P_\mathrm{suc,alg})&=\frac{\sum_{i=1}^r Q_iT_id_i^2}{\prod_i^r k_i\mathcal{P}_{\mathrm{suc}},i}\\
&=\frac{\sum_{i=1}^r Q_i\tau_it_id_i^3}{\prod_i^r k_i\mathcal{P}_{\mathrm{suc},i}},
\end{split}
\end{equation}
where $k_i$ labels the magic states protocol used at each round and $\mathcal{P}_i$ is the probability that round $i$ of distillation will succeed in producing enough magic states to feed the next round. All rounds must succeed for the factory to successfully output $\prod_i^r k_i $ magic states. As discussed $\mathcal{V}$ is a function of several variables, the raw magic state error rate $\epsilon_\mathrm{in}$, the error of gate operations $p_\mathrm{g}$, the total states required $\mathcal{N}$, the protocol chosen $\{k_i\}$, the repetitions allowed in each round $\{t_i\}$ and the probability with which you wish the algorithm to succeed $P_\mathrm{suc,alg}$.

In practice we calculate $\mathcal{V}$ for one, two and three rounds of distillation using combinations of the Bravyi-Haah, Reed Muller and Toffoli protocols with our proposed implementations, with both block and module checking, and a variety of combinations of $t_i$. We then search for the most spacetime efficient method of producing either $T$ or Toffoli magic states, given a $p_\mathrm{g}$, assuming $\epsilon_\mathrm{in}=0.4p_\mathrm{g}$, the number of magic states desired and an arbitrary choice of 90\% for $P_\mathrm{suc,alg}$.

\begin{figure}
\includegraphics[width=\columnwidth]{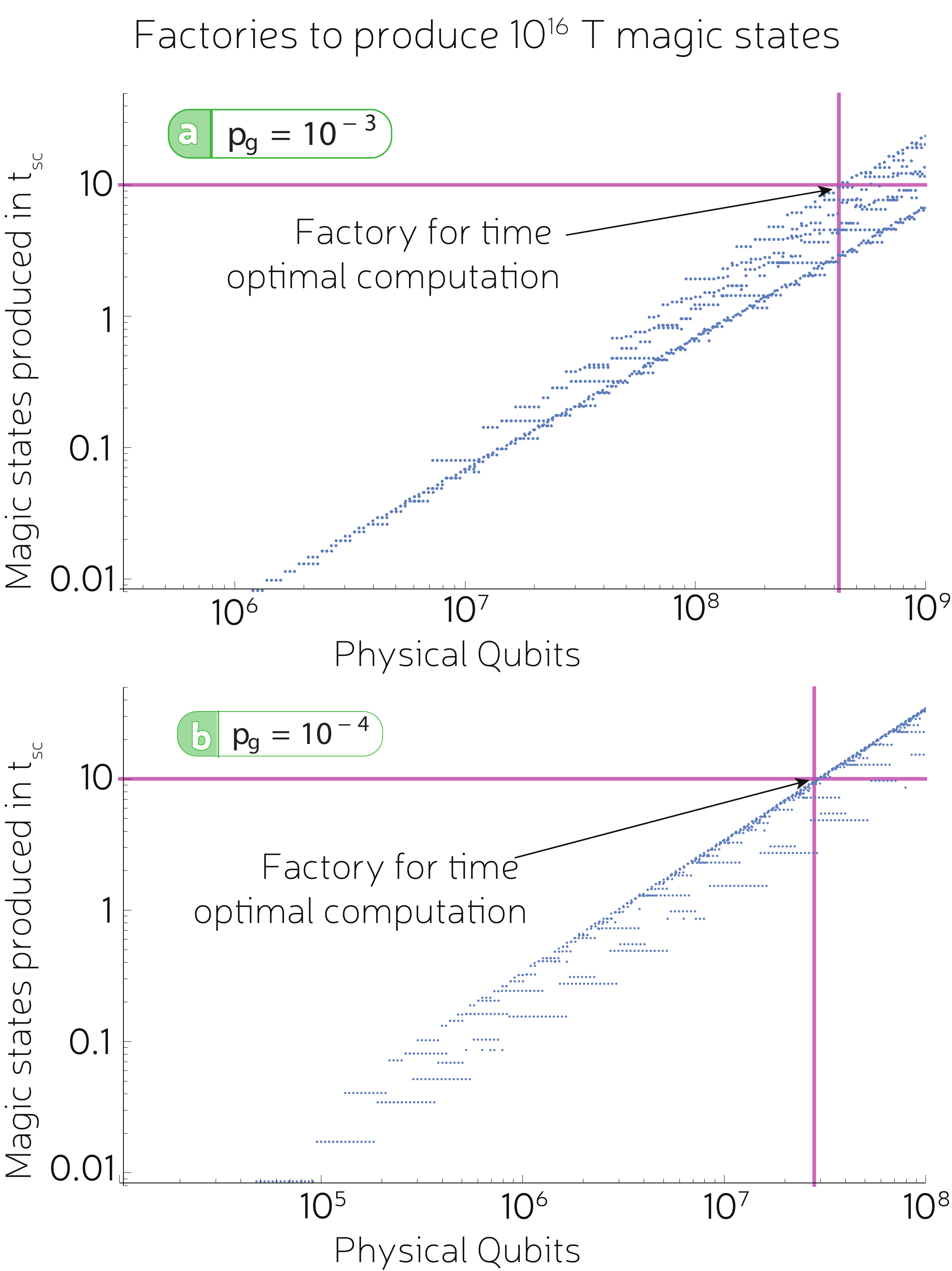}
\caption{To produce \etc{$T$} magic states at a `time-optimal' rate the factory must, on average, produce states at a rate equal to the fastest rate at which they can be used in sequence by an algorithm. In this paper, where we have assumed that $t_{\mathrm{sc}}=10t_{\mathrm{meas/ff}}$, this means 10 magic states must be produced on average every $t_{\mathrm{sc}}$ to keep up with the time optimal implementation of the algorithm. As this figure makes clear, it is indeed possible to produce a given number of magic states faster (slower) than this using more (fewer) qubits. Each data point represents one possible magic state factory given the input error and target number of \textcolor{black}{$10^{16}$ $T$} magic states, only the factories near the boundary between possible and impossible factories are shown. Not shown in the lower-right of each graph are myriad other possible factories of lower rate and higher overhead. As seen clearly in panel (a) doubling the size of the factory can allow one to more than double the rate of magic state production when the larger factory allows the use of the higher $k$ (and therefore higher rate) block codes. This point is less evident in panel (b) where only two rounds of distillation can be sufficient and the higher $k$ block codes are not necessarily optimal. }
\label{timeOptimal}
\end{figure}

The results of these simulations suggest that module checking can provide an improvement of a factor of 3 in the spacetime overhead in certain parameter regimes. We also see that in some regions which, as for the cost analysis, correspond to areas in which there is a transition from $i$ to $i+1$ rounds of Bravyi-Haah and it can be in fact detrimental by a small amount. 


We use the numbers we have generated to estimate the size of a magic state factory required to perform some post-classical factoring tasks using Shor's algorithm. Our results are summarized in Table II, in which we approximate Shor's algorithm as modular exponentiation - as this is by far its most expensive part - and choose the minimum Toffoli gate-count implementation, which has Toffoli-count and -depth equal to $40N^3$ for an $N$ bit number~\cite{Fowler12}. We then determine the minimum possible spacetime overhead per magic state for this task and also the smallest possible factory --- in terms of physical qubits --- that can produce all the magic states necessary while keeping up with a time optimal quantum computation~\cite{Fowler2012}. 

The smallest possible factory is not necessarily the most spacetime efficient factory possible: these factories tend to use larger $k$ blocks which can make the factory formidably vast (cf. Fig.~\ref{FIGtree}), but able to produce more qubits in the same time as lower $k$ protocols. However, these may well produce magic states much faster than required by the fastest possible implementation of the algorithm. Fig.~\ref{timeOptimal} shows that it is possible to use fewer qubits than required by the time optimal implementation and perform your computation at a reduced speed. Equally it is of course possible to increase the size of the factory to produce magic states at a faster rate. In this case though the extra qubit overhead is effectively wasted, unless the computer is performing many calculations in parallel. We find that all the magic states required for a time-optimal factorisation of a 1000 bit number can be produced with a surface code magic state factory of 5.6 million `data qubits', if the infidelity of operations on physical qubits is $10^{-4}$. This is based on the cost of $d^2$ physical qubits to store the information in the rotated lattice surface code~\cite{horsman12}.  Although, for many architectures this number must be doubled to provide ancillas responsible for syndrome extraction.  In this case the physical qubit overhead would be $\sim11$ million.

\begin{figure}
\includegraphics[width=\columnwidth]{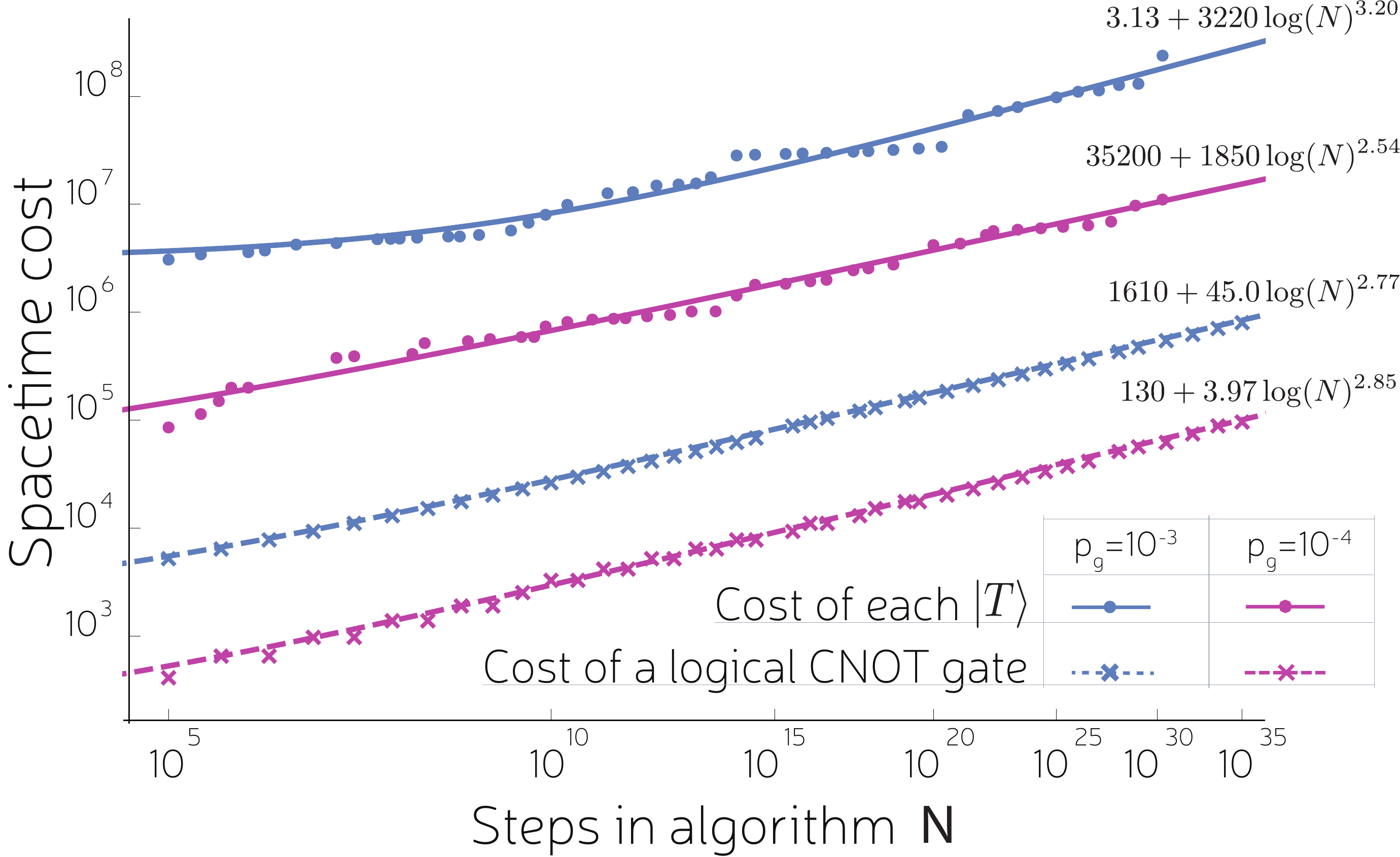} \\
\caption{The scaling of resource cost in qubit-rounds per magic state is not worse than that of the surface code. This corroborates the predictions of Raussendorf~\cite{Raussendorf2007} \textit{et al.} and suggests that the asymptotic overhead scaling $\sim O (\log(N)^3)$ of the surface code is applicable to \textit{universal} fault-tolerant computing with gate counts in the regime of practical interest. Lines are fitted functions of the form $\mathcal{V}(p_{\mathrm{g}},N)=a \log(N)^b+c$}
\label{scaling}
\end{figure}

A summary of the time and space overheads for some example Shor tasks can be found in Table~\ref{ShorTable}. We selected Shor's algorithm as a benchmark as the number of non-Clifford gates required has been well studied and these results have been used in previous analyses. We note that due to the large resource overhead that we have demonstrated, early quantum computers may focus on other problems, particularly in quantum chemistry. A recent analysis of some such problems~\cite{Reiher2016} demonstrates that they are solvable with lower overheads than we find here, albeit that work makes more optimistic assumptions of gate times and fidelities. 

\etc{Of significant interest is how the cost of magic state distillation compares to the surface code overhead.  Raussendorf \textit{et al.}~(see Sec.~6.2 of Ref.~\cite{Raussendorf2007}) were the first to note that using balanced investment in a magic state distillation leads to a constant factor overhead compared to the CNOT gate.  Our numerical results in Fig.~\ref{scaling}  show a ratio between $T$-gate cost and CNOT cost in the range $\sim 150-310$ when $10^{10}<N<10^{30}$.  This ratio is much smaller than estimated by Raussendorf \textit{et al.} who did not make use of Bravyi-Haah distillation routines.  A similar ratio can be extracted from the data tables provided in Ref.~\cite{fowler13}, though the numbers are not directly comparable. For instance, we also count the spacetime volume due to idle qubits, while they wait for distillation circuits to succeed.}

\section{Conclusions}

We proposed the notion of gauge-MSD, which is faster and uses fewer ancillas than previous realisations of Bravyi-Haah magic state distillation.  We further introduced module checking as means to exploit correlations, and found it gave an additional factor $\sim3$ reduction in some parameter regimes. Fowler.~\textit{et al}~\cite{fowler13} considered realising Bravyi-Haah using braiding, and found that Bravyi-Haah only offered a modest $\sim 3$ improvement over the first magic state proposal that used Reed-Muller codes~\cite{BraKit05}.  Therefore, our gains are comparable to, and build upon, other advances in the field.  The work of Bravyi and Haah predicted a much greater improvement because they quantifed cost by the expected conversion efficiency of raw to high-fidelity magic states.  In a fully costed analysis, as we perform here, error correction costs overwhelm and dominate the cost of magic state factories.  We saw that an efficiently designed factory using balanced investment is entirely limited by the surface code cost, and so refinement in distillation protocols can only offer constant factor improvements.

It is likely that this analysis represents an overestimate of the spacetime overhead of the  implementations of the distillation circuits we describe. We have assumed the need for $d$ rounds of surface code measurements after every two-qubit gate. However, it is not clear that this is necessary when performing transversal gates. In the implementation of Bravyi-Haah (see Fig.\ref{BHprotocol}) it may prove feasible to perform $d$ rounds of error correction only after, say, the completion of each of the 4 steps described - reducing the time overhead from $11t_{\mathrm{sc}} d$ to $4t_{\mathrm{sc}} d$~\cite{privateFowler}. Our cost analysis here could thus be further developed by considering the effect on the performance of the underlying surface code if multiple transversal operations are performed between rounds of error correction. However such simulations lie beyond the scope of this paper.

3D gauge color codes~\cite{Bombin13,bombin13b} and other recent ideas do not require magic states.  But they have their own hidden costs. For 3D gauge color codes, spatial overheads scale as $\sim O(\log(N)^3)$ and time overhead as $O(1)$.  Using balanced investment and surface codes we see similar asymptotic scaling of resources. However, current evidence indicates an order of magnitude worse phenomenological threshold for color codes~\cite{Brown16,bravyi15} compared to the phenomenological threshold for the surface code.  Although a full circuit-based threshold has not yet been determined, it is unlikely to challenge that of the surface code due to the higher weight stabilizers required.  This points towards 3D color codes requiring physical error rates of  below $0.1 \%$.   Resource costs are heavily influenced by proximity to threshold, and so 3D color codes seem to require significantly lower physical error rates before they start can compete with surface codes augmented by magic states.  Therefore, with current technology and fidelities, known schemes for avoiding magic states are a false economy.  An additional benefit of the magic state paradigm is that it can also eliminate the additional burden of gate-synthesis costs by preparing exotic magic states~\cite{duclos12,landahl13,duclos15,campbell16,Camp16b,camp16d}. 

\section{Acknowledgments}

Earl Campbell is supported by the EPSRC (grant EP/M024261/1). We thank Mark Howard for comments on the manuscript. We acknowledge support from the NQIT quantum hub in facilitating this collaboration. The authors would like to acknowledge the use of the University of Oxford Advanced Research Computing (ARC) facility in carrying out this work. http://dx.doi.org/10.5281/zenodo.22558.

\appendix

\section{Comparison with braiding}
\label{App::compare}

\etc{Here we discuss how our results compare with prior work on braiding defects in the surface code.  It has been shown that Bravyi-Haah can be realized in constant time~\cite{fowler13} assuming the architecture supports constant time implementations of multi-target CNOT gates (in time $t_{\mathrm{MT-cnot}}$).  This has time cost}
\begin{equation}
 t_{\mathrm{block}}^{(2)} = 12 t_{\mathrm{MT-cnot}} + t_{\mathrm{A}} +  t_{\mathrm{prep}} +  t_{\mathrm{inject}}  +  t_{\mathrm{measure}} ,
\end{equation}
where $t_{\mathrm{inject}}$ is the time to inject a magic state into the circuit, and we can infer $t_{\mathrm{inject}}=t_{\mathrm{cnot}}+t_{\mathrm{measure}}$.  

\etc{In the standard circuit model, multi-target CNOT gates do not take constant time to implement.  In the braiding picture, using ancillary defects, this is possible.  The ``time" cost of braiding 12 such gates is $12 \times 1.25 \times d \times t_{\mathrm{sc}}$.  The total space-time cost was reported as $(96 k+ 216)$ so-called plumbing pieces, which converts into $(\frac{5}{4})^3(96 k+ 216)$ qubit-rounds. It has been recently shown that lattice surgery also supports multi-target CNOT gates in constant time~\cite{herr:latticeSurgery}. Though the Bravyi-Haah protocol was not considered in this setting, one can infer a lattice surgery time cost also scaling with $\sim 12 d $, but with the qubit cost is not currently unknown.}

The above discussion implies \etc{a modest space-time saving of using gauge-MSD in a distributed architecture rather than braiding in a nearest neighbour picture. We remark that gate times and qubit expense will vary on a much greater scale between different hardware platforms.  In particular, the long-range gates of distributed schemes are often much slower, with photonics protocols impeded by photon loss and the potential need for entanglement purification~\cite{BK01a,JTSL02a,ODH01a,Moehring07,Camp08,Camp2010review,Hanson13,Nickerson14}.}

\section{Formal tools}
\label{App:Formal}

\subsubsection{Stabilizer Bravyi-Haah codes}

Let us begin with some basic concepts from code theory and related notation. Stabilizer codes are subspaces described by an abelian group called the stabilizer $\mathcal{S}$, which is a subgroup of the Pauli group.  The projector on to the codespace is $\Pi \propto \sum_{s \in \mathcal{S}} s$, so that $s \Pi =\Pi$ for all $s \in \mathcal{S}$.  There always exists a minimal set of operators $\{ S_1, S_2, ... S_m \}$ that generate the group, which we denote by $\mathcal{S}= \langle S_1, S_2, ... S_m \rangle $.  For the CSS codes, these generators can be chosen so that they are all either X-type or Z-type, as we define shortly. If $g$ is a binary vector of length $n$ we use $Z[g]=\otimes_{j : [g]_j= 1}^n Z_j$ and similarly $X[g]=\otimes_{j : [g]_j= 1}^n X_j$. We say $Z[g]$ are $Z$-type operators, and $X[g]$ are $X$-type operators.  Therefore, a CSS code has generators $\mathcal{S}=\langle Z[f_1], \ldots, Z[f_a],   X[g_1], \ldots, X[g_b] \rangle$.  Commutation of $X[g]$ and $Z[f]$ is equivalent to $(f,g)=0$ where we use the inner product $(f,g):= f \cdot g \pmod{2}$. 

Bravyi and Haah introduced the notion of a $G$ matrix which is a binary matrix composed of two submatrices $G_1$ and $G_0$. We label the rows of $G$ as $\{ g_1, \ldots , g_b, g_{b+1}, \ldots g_m \}$ where the rows $\{ g_1, \ldots , g_b \}$ belong to $G_0$ and the rows $\{ g_{b+1}, \ldots , g_m \}$ belong to $G_1$.  These matrices define a CSS code as follows.  The rows of $G_0$ are some set $ \{ g_1, \ldots, g_b \} $ that specify the $X$-type generators $\{ X[g_1], \ldots, X[g_b] \}$. The $Z$-type generators are given less explicitly.  Denote $\mathcal{G}^{\perp}$ to be the binary vector space orthogonal to both $G_0$ and $G_1$, that is $\mathcal{G}^{\perp} := \{ g : (f,g) = 0 ;  \forall f \in G_0, G_1  \}$.  Note that Bravyi-Haah required that $G_0 \subset \mathcal{G}^\perp$.  We define $G^{\perp}$ to be some (minimal) matrix with rows $\{ f_1, \ldots f_a \}$ that generate the group $\mathcal{G}^{\perp}$ under row-wise modular addition.  This defines the $Z$-type generators $\{ Z[f_1], \ldots, Z[f_b] \}$.  Note that there exist many different choices for $G^{\perp}$, which all result in the same CSS code with $Z$-type generators $\{ Z[f_1], \ldots, Z[f_a] \}$.  

This completes the description of the stabilizer codespace, though we also need to know how information is stored within the subspace.  We have that the $X$ operator for the $k^{\mathrm{th}}$ logical qubit is $X[g_k]$ where $g_k$ is a row of $G_1$, and so $b+1 \leq k \leq m$.  These are representatives of the logical operators, with equivalent logical operators differing by only a stabilizer.  For Bravyi-Haah protocols all rows in $G_1$ have odd weight, so we may also take that the $Z$ operator for the $k^{\mathrm{th}}$ logical qubit \textcolor{black}{as} $Z[g_k]$ where $g_k$ is the $k^{\mathrm{th}}$ row of $G_1$. 

\subsubsection{Subsystem Bravyi-Haah codes}

Given a $G$-matrix we define a subsystem code with stabilizer $\tilde{\mathcal{S}}:=\langle Z[g_1], \ldots, Z[g_b],  X[g_1], \ldots, X[g_b] \rangle$ where $\{ g_1, \ldots, g_b \}$ are the rows of $G_0$.  Notice that now there are equal numbers of $X$ and \etc{$Z$} stabilizers and they both correspond to the rows of $G_0$.  Bravyi and Haah considered a class of matrices obeying triorthgonality conditions, which require that $G_0 \subset \mathcal{G}^\perp$.  Therefore, we have that $\tilde{\mathcal{S}} \subset \mathcal{S}$, with the subsystem code having strictly fewer stabilizers than the original Bravyi-Haah code.  We denote the subsystem projector as $\tilde{\Pi}\propto \sum_{s \in \tilde{\mathcal{S}}} s$.  We further take the logical operator for the subsystem code to be identical to those of the original Bravyi-Haah code.  This leaves some degrees of freedom as neither stabilizers nor logical operators.  We define the gauge group $\mathcal{S}_g:= \langle Z[f_1], \ldots, Z[f_a],   X[f_1], \ldots, X[f_a] \rangle $ where $\{ f_1, \ldots f_a \}$ are rows of $G^\perp$.  Notice that $\mathcal{S}_g$ contains $\mathcal{S}$, by virtue of $G_0 \subset \mathcal{G}^\perp$. 

Let us recap.  Our subsystem code is defined by its stabilizer $\tilde{\mathcal{S}}$ and gauge group $\mathcal{S}_g$, whereas the original Bravyi-Haah code has stabilizer $\mathcal{S}$, and these groups satisfy $\tilde{\mathcal{S}} \subset \mathcal{S} \subset \mathcal{S}_g$. However, $\mathcal{S}_g$ is inflated in size compared to $\mathcal{S}$ and is no longer abelian.  Furthermore, one can verify that $\mathcal{S}_g$ does not contain any logical operators as follows. First, note that Bravyi-Haah use triorthogonal (also called triply even) matrices where for any $f,g \in G$ we have that $(f,g)=1$ if and only if $f=g$ and $f \in G_1$. 

As logical operators we take $X[l]$ and $Z[l]$ for each $l$ in $G_1$.   From the triorthogonality of $G$ we see that $X[l]$ and $Z[l]$ anticommute, but $X[l]$ and $Z[l']$ commute when $l \neq l'$.  To properly describe a subsystem code, where measuring the gauge operators does not damage the logical qubits, we require that the logical operators are not elements of the gauge group $\mathcal{S}_g$. Recall the gauge group is defined by vectors that reside in the dual code $\mathcal{G}^\perp$.  Therefore, every gauge operator must have vanishing inner product with every row in $\mathcal{G}$.  However, $l \in \mathcal{G}$, and $(l,l)=1$, so $l$ is not in the dual space and the logical operators are not gauge operators. This completes our proof that the logical operators indeed lie outside the gauge group.

\section{Realising the Bravyi-Haah protocols}
\label{SubSystem}
  
\textcolor{black}{There are many routes to realising magic state distillation. Assuming perfect Clifford operations, different realisations suppress errors equally, but differ in terms of temporal depth and required ancillas.  Many of these potential realisations have only been sketched, without a complete assessment of resources involved.  Here we introduce a new method particularly suitable for architectures implementing logical gates via transversal operations or lattice surgery~\cite{horsman12}. Conceptually, we are inspired by notions of subsystem codes and gauge fixing techniques, and so call our approach gauge-MSD.  We consider only even $k$ with $k \in \{ 2,6,\ldots, 4m+2, \ldots \}$ as then the Bravyi-Haah codes have transversal $T$ gates.  For  $k \in \{ 0,4,8,\ldots, 4m, \ldots \}$,  the Bravyi-Haah codes have transversal $T$ gates, only up to a nonlocal Clifford correction.}

\subsection{Outline of protocol}

\textcolor{black}{Here we present an outline of gauge-MSD, with details of how to realise multiqubit Pauli measurements postponed until the next section.  First we specify some notation. Refer back to App.~\ref{App:Formal} for definitions of the $G$-matrix and $G^\perp$-matrix.  Let $R$ be a binary matrix such that $G^\perp \cdot R = \id \pmod{2}$, which is ensured to exist by virtue that $G^\perp$ is full rank.  Furthermore, let $M$ be a binary matrix such that $M \cdot G^\perp = G_0 \pmod{2}$, which must exist since $G_0$ is in the span of $G^\perp$.  Explicit examples of $G^\perp$, $R$ and $M$ will be given in the next section.  Measurement outcomes will be recorded in binary: $0$ for $``+1"$ eigenvalues and $1$ for $``-1"$ eigenvalues.  Let $\mathcal{O}$, $\mathcal{X}$, $\mathcal{Z}$ be disjoint sets
\begin{align}
	\mathcal{O} &= \{ 6+3j | j =1,2,\ldots k\} , \\ \nonumber
	\mathcal{X} &= \{1,2,3\} , \\ \nonumber
	\mathcal{Z} &= \{4,5,6,7,8,7+3j,8+3j |  j= 1,2,\ldots,k \} .
\end{align}
Associated with these sets are binary matrices that allow us to compute Pauli corrections, $H_Z$ is a $k$-by-$|\mathcal{X}|$ matrix and $H_X$ is a $k$-by-$|\mathcal{Z}|$ matrix as follows
\begin{align}
H_Z & = \begin{blockarray}{cccccccccccc}
4 & 5 & 6 & 7 & 8  & 10 & 11 & 13 & 14 & 16 & 17 & \ldots  \\
\begin{block}{(cccccccccccc)}
  0 & 1 & 1 & 1 & 1 & 1 & 1 & 0 & 0 & 0 & 0 & \\
  0 & 1 & 1 & 1 & 1 & 0 & 0 & 1 & 1 & 0 & 0 & \ldots \\
  0 & 1 & 1 & 1 & 1 & 0 & 0 & 0 & 0 & 1 & 1 & \\
   & &  &  &  &  \svdots &  &  &  \svdots & &  & \\
\end{block} 
\end{blockarray} \\
H_X & = \begin{blockarray}{ccc}
1 & 2 & 3 \\ 
\begin{block}{(ccc)}
	1 & 1 & 0 \\
	1 & 1 & 0 \\
	1 & 1 & 0 \\
   & \svdots & \\
\end{block}
\end{blockarray} 
\end{align}
where the numbers above the columns correspond to the elements in sets $\mathcal{Z}$ and $\mathcal{X}$.  Lastly, we will also make use of a $k$-by-$\mathrm{dim}(G^\perp)$ matrix denoted $Q$ that satisfies $G_1 = W + Q G^\perp$ where $W$ has $W_{j,6j+3}=1$, $W_{j,1}=W_{j,2}=1$ and zero everywhere else. An example $Q$ is given in the next section.}

\textcolor{black}{We now state the protocol. 
\begin{enumerate}
    \item Measure all $Z[f_k]$ where $f_k$ is the $k^{\mathrm{th}}$ row of $G^\perp$, recording outcomes as $\mu = (\mu_1 , \mu_2, \ldots , \mu_k )$;
    \item Apply $A[w]$ where $w=R\mu \pmod{2}$;
    \item Measure all $X[f_k]$ where $f_k$ is the $k^{\mathrm{th}}$ row of $G^\perp$, recording outcome as $\gamma = ( \gamma_1 , \gamma_2, \ldots , \gamma_k)$;  
	\item Declare SUCCESS if $M \gamma = (0,\ldots, 0) \pmod{2}$, and FAILURE otherwise;
	\item Measure qubits in set $\mathcal{X}$ in the $X$ basis, recording outcomes as $m_X$; Simultaneously, measure qubits in set $\mathcal{Z}$ in the $Z$ basis, recording outcomes as $m_Z$.  
    \item Qubits in $\mathcal{X}$ and $\mathcal{Z}$ are discarded, while qubits in set $\mathcal{O}$ are retained as output as qubits $(1,2,\ldots,k)$;
	\item Apply Pauli corrections $X[H_Z m_Z] Z[ H_X m_X + Q \gamma]$, or update Pauli frame accordingly.
\end{enumerate}
After steps 1-2, the state is deterministically projected by $\Pi_Z \propto \sum_{s \in \mathcal{S}_Z} s$ where $\mathcal{S}_Z := \langle Z[f_1], \ldots, Z[f_a] \rangle$, exactly as in the original Bravyi-Haah protocol.  After step 3, the system is an eigenstate of $(-1)^{\gamma_k}X[f_k]$ and we can associate some projector with $\Pi_{X, \gamma}$ with this process.  The postselection in step 4 ensures the system is an eigenstate of $+X[g_k]$ where $g_k$ is the $k^{\mathrm{th}}$ row of $M \cdot G^\perp $.  Since we defined $M$ such that $M \cdot G^\perp = G_0$, we have that $g_k$ are the rows of $G_0$.  It follows that $\Pi_{X, \gamma} = \Pi_{X} \Pi_{X, \gamma}$ where $\Pi_{X}$ is the projector for the  $X$ stabilizer of the Bravyi-Haah stabilizer code.  Combining, steps 1-4 we have the projections 
\begin{equation}
	\Pi_{X, \gamma} \Pi_{Z} = \Pi_{X, \gamma} \Pi_{X}  \Pi_{Z} =   \Pi_{X, \gamma} \Pi ,
\end{equation} 
where $\Pi = \Pi_{X}  \Pi_{Z} $ is the full  projector onto the Bravyi-Haah stabilizer codespace.}

\textcolor{black}{We have obtained the desired $\Pi$ projection required by the Bravyi-Haah protocol.  But we have picked up an additional $\Pi_{X,\gamma}$. This additional projection results from measuring some the gauge operators of the subsystem variant of Bravyi-Haah. Thus the logically encoded qubits are unharmed but there has been a change to the gauge degrees of freedom. }

\textcolor{black}{In step 5, we perform single qubit measurements to isolate the $k$ logical qubits from the $n$ qubits in the subsystem code.  Recall that in the original presentation of Bravyi Haah we have logical operators $Z[l_j]$ and $X[l_j]$ for the $j^{\mathrm{th}}$ logical qubit, where $l_j$ is the $j^{\mathrm{th}}$ row of $G_1$.  The measurements localise these logical operators onto single qubits, up to a Pauli correction depending on the measurement outcomes.  That is, the measurements cause the state to become stabilized by some new $\pm Z[w]$ operators, and every logical $Z[l_j]$ can be multiplied by these operators to obtain a single qubit $\pm Z$ operator acting on the $(6+3j)^{\mathrm{th}}$ qubit.  It is straightforward to verify that the $\pm$ sign is corrected to $+$ by the Pauli correction $X[H_Z m_Z]$. To show that $X$ logical operators are also localized on a single output qubit, we first multiply $X[l_j]$ by $X$-type gauge operators until it is acts on qubits $2$,$3$ and $(6+3j)$.  Measuring qubits 1,2 and 3 completes the localisation of $X[l_j]$ onto the $(6+3j)^{\mathrm{th}}$ qubit.  The required Pauli operator is now $X[H_Z m_Z + Q \gamma]$, where now there is also some dependence on the eigenvalues of gauge operators obtained in step 3.}

\subsection{Implementing Pauli measurements in minimum depth}

\begin{figure}
\includegraphics[width=\columnwidth]{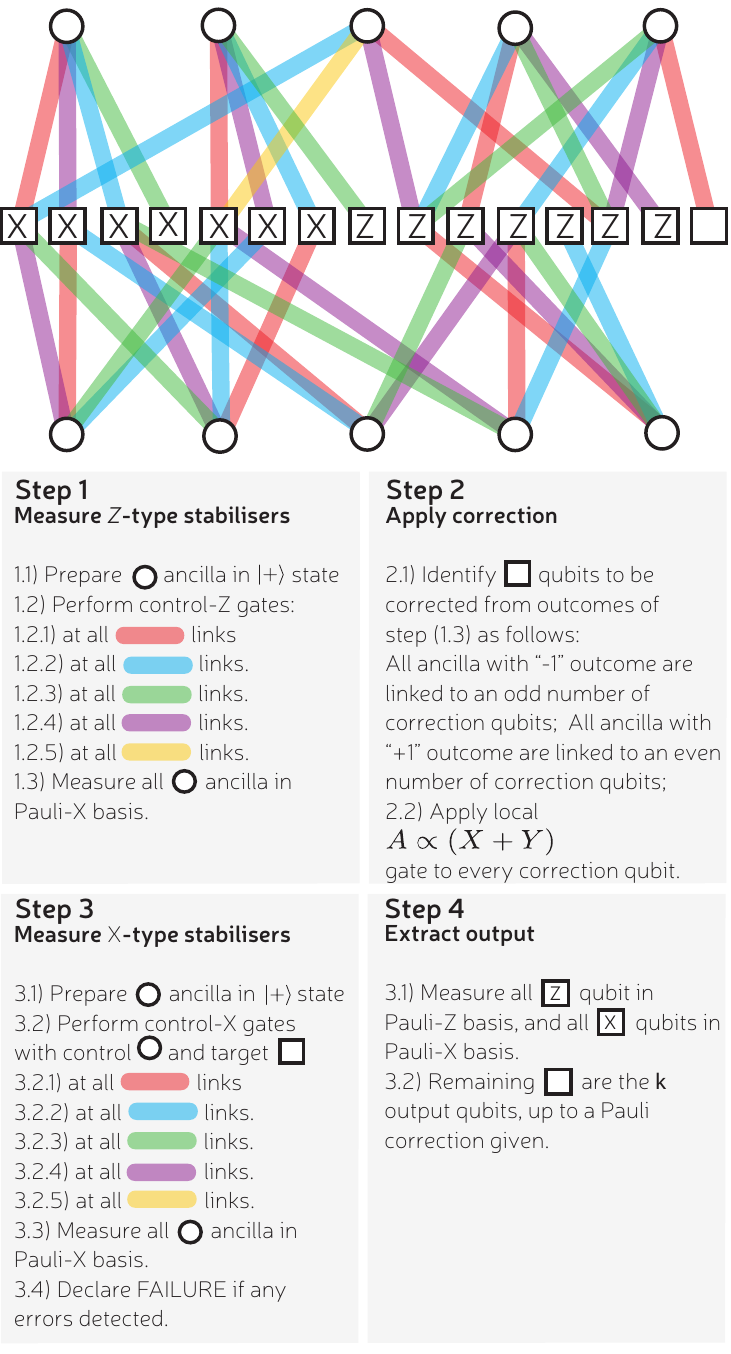}[t]
\caption{Squares represent magic states to be distilled, and circles are ancillas used to impleme\textcolor{black}nt measurements in Reed-Muller.  Edges show required hardware connections and associated required control-phase gates.  Edges show time ordering of control phase gates, e.g. red, blue, purple green and gold, demonstrating the required entanglement can be established in 5 time steps. Sometimes we call this the graph representing $G_z^{\text{RM}}$.}
\label{reasliseRM}
\end{figure}

Implementing the protocol requires a set of $Z$ measurements, followed by a set of $X$ measurements.  In the original standard implementation the $X$ measurements involved many qubits and so \textcolor{black}{were} difficult to implement.  However, the previous subsection shows that the difficult $X$ measurements can be replaced with lower weight measurements mirroring the $Z$ measurements performed.  \textcolor{black}{The complexity of these measurements depends on the row weights of $G^\perp$.  The matrix $G^\perp$ must generate the space $\mathcal{G}^\perp$, but there is some freedom in how we choose the generating rows.  In the work of Bravyi-Haah, $G^\perp$ was only implicitly defined (as the dual of $G$), leaving it unclear how much time it would take to implement the required measurements. It is desirable that $G^\perp$ is as sparse as possible to minimise the resource overheads. Therefore, we wish to find a very sparse $G^\perp$.  We found a family of $G^\perp$ matrices where all rows, except one, are weight 4 and the single exception has weight $k+2$.  We present the exact form of this $G^\perp$ for $k=2$, along with associated $R$, $M$ and $Q$ matrices used in the previous section.
\begin{align}
    G^\perp & = \left( \begin{array}{cccccccc|ccc|ccc}
 1 & 0 & 0 & 1 & 0 & 1 & 1 & 0 & 0 & 0 & 0 & 0 & 0 & 0 \\
 0 & 1 & 0 & 1 & 1 & 0 & 1 & 0 & 0 & 0 & 0 & 0 & 0 & 0 \\
 0 & 0 & 1 & 1 & 1 & 1 & 0 & 0 & 0 & 0 & 0 & 0 & 0 & 0 \\
 0 & 0 & 0 & 0 & 1 & 0 & 0 & 1 & 1 & 1 & 0 & 0 & 0 & 0 \\    
 0 & 0 & 0 & 0 & 0 & 1 & 0 & 1 & 1 & 0 & 1 & 0 & 0 & 0 \\ 
 0 & 0 & 0 & 0 & 0 & 0 & 1 & 1 & 0 & 1 & 1 & 0 & 0 & 0 \\ 
 0 & 0 & 0 & 0 & 0 & 0 & 0 & 0 & 1 & 1 & 0 & 1 & 1 & 0 \\ 
 0 & 0 & 0 & 0 & 0 & 0 & 0 & 0 & 0 & 1 & 1 & 0 & 1 & 1 \\         
 0 & 0 & 1 & 0 & 0 & 0 & 1 & 0 & 1 & 0 & 0 & 1 & 0 & 0
     \end{array}
  \right) ,  \\ \nonumber
    R & = \left( \begin{array}{ccccccccc}
            1& 0& 0& 0& 1& 1& 0& 0& 0\\
            0& 1& 0& 1& 0& 1& 0& 0& 0\\
            0& 0& 1& 1& 1& 0& 0& 0& 0\\
            0& 0& 0& 0& 0& 0& 0& 0& 0\\
            0& 0& 0& 1& 0& 0& 0& 0& 0\\
            0& 0& 0& 0& 1& 0& 0& 0& 0\\
            0& 0& 0& 0& 0& 1& 0& 0& 0\\
            0& 0& 0& 0& 0& 0& 0& 0& 0\\
            0& 0& 0& 0& 0& 0& 0& 0& 0\\
            0& 0& 0& 0& 0& 0& 0& 0& 0\\
            0& 0& 0& 0& 0& 0& 0& 0& 0\\
            0& 0& 1& 1& 1& 1& 0& 0& 1\\
            0& 0& 1& 1& 1& 1& 1& 0& 1\\
            0& 0& 1& 1& 1& 1& 1& 1& 1\\      
     \end{array} \right) , \\ \nonumber
   Q & = \left( \begin{array}{ccccccccc}
            1 & 1& 0& 0& 0& 1& 0& 0& 0\\
            1 & 1& 0& 0& 0& 1& 0& 1& 0\\
     \end{array} \right), \\ \nonumber
M & = \left( \begin{array}{ccccccccc}
0& 1& 0& 1& 1& 1& 1& 1& 0 \\ 
0& 0& 1& 1& 1& 1& 0& 1& 0 \\ 
1& 1& 1& 1& 1& 1& 0& 0& 0 \\
\end{array} \right)
\end{align} }

\textcolor{black}{A Mathematica script for generating $G^\perp$ for any $k$ is provided in the supplementary material, which also verifies that $G^\perp$ is full rank and dual to $G$ and.  As one further example, we find for $k=6$ that}

\begin{widetext}
\begin{equation}
    G^\perp = \left( \begin{array}{cccccccc|ccc|ccc|ccc|ccc|ccc|ccc}
1& 0& 0& 1& 0& 1& 1& 0& 0& 0& 0& 0& 0& 0& 0& 0& 0& 0& 0& 0& 0& 0& 0& 0& 0& 0\\ 
0& 1& 0& 1& 1& 0& 1& 0& 0& 0& 0& 0& 0& 0& 0& 0& 0& 0& 0& 0& 0& 0& 0& 0& 0& 0\\ 
0& 0& 1& 1& 1& 1& 0& 0& 0& 0& 0& 0& 0& 0& 0& 0& 0& 0& 0& 0& 0& 0& 0& 0& 0& 0\\ 
0& 0& 0& 0& 1& 0& 0& 1& 1& 1& 0& 0& 0& 0& 0& 0& 0& 0& 0& 0& 0& 0& 0& 0& 0& 0\\ 
0& 0& 0& 0& 0& 1& 0& 1& 1& 0& 1& 0& 0& 0& 0& 0& 0& 0& 0& 0& 0& 0& 0& 0& 0& 0\\ 
0& 0& 0& 0& 0& 0& 1& 1& 0& 1& 1& 0& 0& 0& 0& 0& 0& 0& 0& 0& 0& 0& 0& 0& 0& 0\\ 
0& 0& 0& 0& 0& 0& 0& 0& 1& 1& 0& 1& 1& 0& 0& 0& 0& 0& 0& 0& 0& 0& 0& 0& 0& 0\\
0& 0& 0& 0& 0& 0& 0& 0& 0& 1& 1& 0& 1& 1& 0& 0& 0& 0& 0& 0& 0& 0& 0& 0& 0& 0\\ 
0& 0& 0& 0& 0& 0& 0& 0& 0& 0& 0& 1& 1& 0& 1& 1& 0& 0& 0& 0& 0& 0& 0& 0& 0& 0\\ 
0& 0& 0& 0& 0& 0& 0& 0& 0& 0& 0& 0& 1& 1& 0& 1& 1& 0& 0& 0& 0& 0& 0& 0& 0& 0\\ 
0& 0& 0& 0& 0& 0& 0& 0& 0& 0& 0& 0& 0& 0& 1& 1& 0& 1& 1& 0& 0& 0& 0& 0& 0& 0\\ 
0& 0& 0& 0& 0& 0& 0& 0& 0& 0& 0& 0& 0& 0& 0& 1& 1& 0& 1& 1& 0& 0& 0& 0& 0& 0\\ 
0& 0& 0& 0& 0& 0& 0& 0& 0& 0& 0& 0& 0& 0& 0& 0& 0& 1& 1& 0& 1& 1& 0& 0& 0& 0\\ 
0& 0& 0& 0& 0& 0& 0& 0& 0& 0& 0& 0& 0& 0& 0& 0& 0& 0& 1& 1& 0& 1& 1& 0& 0& 0\\ 
0& 0& 0& 0& 0& 0& 0& 0& 0& 0& 0& 0& 0& 0& 0& 0& 0& 0& 0& 0& 1& 1& 0& 1& 1& 0\\ 
0& 0& 0& 0& 0& 0& 0& 0& 0& 0& 0& 0& 0& 0& 0& 0& 0& 0& 0& 0& 0& 1& 1& 0& 1& 1\\ 
0& 0& 1& 0& 0& 0& 1& 0& 1& 0& 0& 1& 0& 0& 1& 0& 0& 1& 0& 0& 1& 0& 0& 1& 0& 0
     \end{array} \right) , 
\end{equation}
\end{widetext}
\textcolor{black}{Notice that the bottom row has weight exceeding four.  The measurements corresponding to weight four rows can be implemented with each measurement using a single ancilla in the $\ket{+}$ state and four entangling gates (control-$Z$ or control-$X$ depending on the measurements).  Therefore, for these measurements it is possible that all these gates can be realized in four time steps, while respecting that a qubit can only be involved in a single entangling gate at a time.  Unfortunately, there is a single row of $G^\perp$ with weight $k+2$, and so using a single ancilla to perform this measurement would result a growing time cost with $k$.  Therefore, for this single measurement we make use of a cat state $\ket{0}^{\otimes k+2}+\ket{1}^{\otimes k+2}$ so that the entangling gates can be performed in parallel.  The cat state itself is constructed by merge operators on $\ket{+}^{\otimes k+2}$ qubits. The merge operations projecting onto $\kb{00}{00}+\kb{11}{11}$ or $\kb{01}{01}+\kb{10}{10}$ subspaces and so commute with the control gates and so the cat state can be built concurrently with the control gates.  This opens the possibility of realising each round of measurements in 4 times steps, but depends on whether the entangling gates can be scheduled in an economical manner. The scheduling problem is equivalent to a graph coloring problem and we find that it can be solved in 4 time steps (e.g. using 4 colors) as in Fig.~\ref{BHprotocol}.  We independently confirmed this using an automated solver of the edge colorability problem for $k$ up to 40, see supplementary Mathematica script for details. }

\section{Reed-Muller connectivity}

The usual form of the 15-qubit punctured Reed-Muller code the $G$ matrix is

\begin{equation}
  \resizebox{0.95\hsize}{!}{$     G_{\text{RM}} = \left( \begin{array}{ccccccccccccccc}
      1 & 1 & 1 & 1 & 1 & 1 & 1 & 1 & 0 & 0 & 0 & 0 & 0 & 0 & 0 \\
      1 & 1 & 1 & 1 & 0 & 0 & 0 & 0 & 1 & 1 & 1 & 1 & 0 & 0 & 0 \\
      1 & 1 & 0 & 0 & 1 & 1 & 0 & 0 & 1 & 1 & 0 & 0 & 1 & 1 & 0 \\
      1 & 0 & 1 & 0 & 1 & 0 & 1 & 0 & 1 & 0 & 1 & 0 & 1 & 0 & 1 \\  \hline
	  1 & 1 & 1 & 1 & 1 & 1 & 1 & 1 & 1 & 1 & 1 & 1 & 1 & 1 & 1 
     \end{array}
  \right)$}
\end{equation}
For our purposes an efficient choice of dual is
\begin{equation}
  \resizebox{0.95\hsize}{!}{$   G_{\text{RM}}^{\perp} = \left( \begin{array}{ccccccccccccccc}
    0 & 0 & 0 & 0 & 1 & 1 & 1 & 1 & 0 & 0 & 0 & 0 & 0 & 0 & 0 \\
    0 & 0 & 0 & 0 & 0 & 0 & 0 & 0 & 1 & 1 & 1 & 1 & 0 & 0 & 0 \\
    0 & 0 & 0 & 0 & 0 & 0 & 0 & 0 & 1 & 1 & 0 & 0 & 1 & 1 & 0 \\
    0 & 0 & 0 & 0 & 0 & 0 & 0 & 0 & 1 & 0 & 1 & 0 & 1 & 0 & 1 \\ 
    1 & 1 & 1 & 1 & 0 & 0 & 0 & 0 & 0 & 0 & 0 & 0 & 0 & 0 & 0 \\
    1 & 1 & 0 & 0 & 1 & 1 & 0 & 0 & 0 & 0 & 0 & 0 & 0 & 0 & 0 \\
    1 & 0 & 1 & 0 & 1 & 0 & 1 & 0 & 0 & 0 & 0 & 0 & 0 & 0 & 0 \\
    0 & 1 & 1 & 0 & 0 & 0 & 0 & 0 & 0 & 1 & 1 & 0 & 0 & 0 & 0 \\ 
	0 & 0 & 1 & 0 & 1 & 0 & 0 & 0 & 0 & 0 & 1 & 0 & 1 & 0 & 0 \\
    1 & 0 & 0 & 0 & 1 & 0 & 0 & 0 & 1 & 0 & 0 & 0 & 1 & 0 & 0 \\     
     \end{array}
  \right)$}
\end{equation}
which has a maximum row weight of 4 and a max column weight 5. There is a 5 colorable graph associated with this matrix  shown in Fig.~\ref{reasliseRM}, similar to what we found for the Bravyi-Haah protocol.  It is well known that the Reed-Muller code can also be viewed as a subsystem code, and so the Pauli-$X$ measurements can clone the pattern of the Pauli-$Z$ measurements.  This approach yields the realisation shown in  Fig.~\ref{reasliseRM}.

\section{Toffoli protocol}
\label{Toff}

\begin{figure}
\includegraphics[width=\columnwidth]{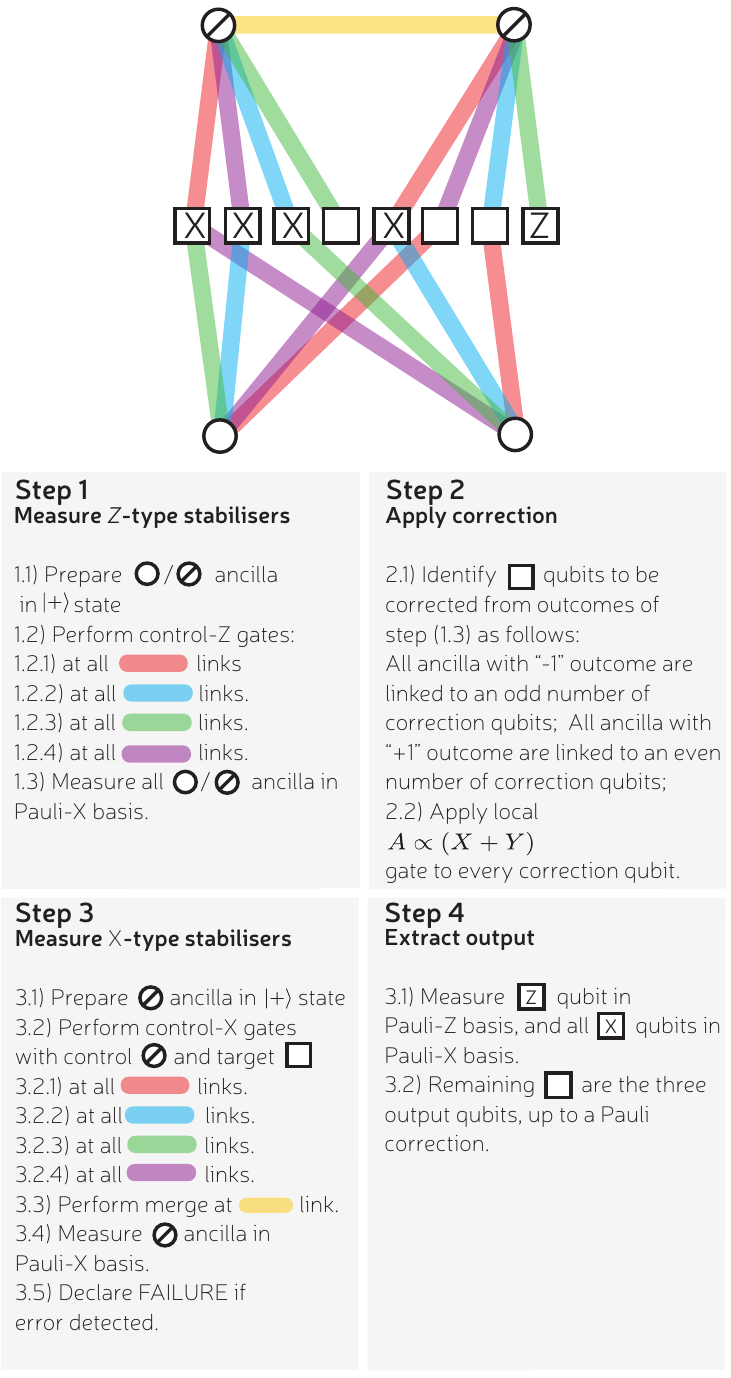}
\caption{Squares represent magic states to be distilled, and circles are ancillas used to implement the stabilizer measurements which project the T-states into the Toffoli state.  As before edges show required hardware connections and associated required control-phase gates.  Edges show time ordering of control phase gates, e.g. red, blue, purple green and gold, demonstrating the required entanglement can be established in 4 time steps for the Z stabilizers and 5 time steps for the single X stabilizer, which utilises just \textcolor{black}{two} of the four ancillas. Sometimes we call this the graph representing $G_z^{Toff}$.}
\label{Toff_Prot}
\end{figure}

The Toff protocol converts 8 $T$-states into one Toffoli state
\begin{equation}
	\ket{\mathrm{Tof}} = \frac{1}{\sqrt{8}} \sum_{a,b,c \in \{ 0,1\} } (-1)^{abc} \ket{a,b,c}.	
\end{equation}
It has previously been described in the circuit picture with its performance found by brute force counting of errors~\cite{eastin13,jones13b}.  Here we consider an equivalent protocol (with the same performance when using block checking) but with the $G$ matrix formalism as the Bravyi-Haah protocols.  The $G$ matrix achieving this is simply
\begin{equation}
	G_{\mathrm{Tof}} = \left( \begin{array}{cccccccc}
	1 & 1 & 1 & 1 & 0 & 0 & 0 & 0 \\
	1 & 1 & 0 & 0 & 1 & 1 & 0 & 0 \\
	1 & 0 & 1 & 0 & 1 & 0 & 1 & 0 \\ \hline
	1 & 1 & 1 & 1 & 1 & 1 & 1 & 1
 \end{array} \right),
\end{equation}
where proof that this distills was given in Ref.~\cite{Camp16b,camp16d}.  Here we show in Fig.~\ref{Toff_Prot} how to convert this abstract protocol to a realisation with low spacetime overhead.

\section{Proof of error tracking}
\label{proof}

When iterating distillation protocols we have a collection of inputs, which may have been outputs of \textcolor{black}{an} earlier round.  We illustrate the structure in Fig.~\ref{FIGtree}, where we introduce the terminology of branches and modules.  Our proof proceeds by considering how an individual module maps probability distributions over its inputs to a probability distribution over outputs.  If a level-$l$ module uses many $n_l \rightarrow k_l$ blocks, then it requires $n_l$ branches of inputs from earlier rounds.  If each branch carries $N_l$ qubits, then we need $N_l$ instances of the $n_l \rightarrow k_l$ protocol so that the whole module maps $N_ln_l \rightarrow N_lk_l$ qubits.  The size of $N_l$ equals the product of the $k_l$ values for all earlier rounds, which can be verified by simply counting back through the tree.  

We use $x^{(j)}\in \mathbb{Z}_2^N$ to denote the error distribution of the input contribution from the $j^{\mathrm{th}}$ branch, and use probability $\Pr (x^{(j)})$ for the probability of this event.  For now we take this probability as given and note only that for a block that quadratically suppresses noise, we have that $\Pr(x^{(j)}\neq 0)$ is to first order proportional to $\epsilon^{2^l}$
after $l$ rounds of distillation.  If a single block of a $n_l \rightarrow k_l$ protocol is described by a matrix $G$, then $N$ copies within a module are described by $\mathcal{G}=G \otimes \id_N $ where $\otimes$ is the tensor product and $\id_N $ is the identity matrix of dimension $N$.  On the first round, $N=1$ and so we simply have $\mathcal{G}=G$. Before proceeding we must clarify how this tensor product structure relates to the input strings $x$. We define $\delta^{(j)}$ as a length $n_l$ binary vector with entries: $1$ in the $j^{\mathrm{th}}$ location, and 0 everywhere else.  It follows that $x=\sum_j (\delta^{(j)} \otimes x^{(j)})$, so that $\mathcal{G}x=\sum_j (G \delta^{(j)} \otimes x^{(j)})$.  This ordering ensures that no two qubits from the same branch collide into the same block, which must be prevented because of correlations within branches.  We call this the canonical ordering and it is assumed throughout. Other orderings exist that prevent such collisions, such as an arbitrary permutation of qubits within any branch.  Potentially such permutations could perform better or worse than the canonical choice, leaving room for further optimisation.  We continue with the canonical choice as it is particularly natural and amenable to analysis. 

After a module passes module checking, the output branch has errors distributed as
\begin{equation}
    \Pr_l( y ) = \frac{1}{\ps{l}}  \sum_{ \{ x :  \mathcal{G}_0 x = 0 , \mathcal{G}_1 x = y   \} } \Pr( x ) ,  
\end{equation}
where the denominator is a normalisation constant accounting for the success probability 
\begin{equation}
    \ps{l} = \sum_{y}  \sum_{ \{ x :  \mathcal{G}_0 x = 0, \mathcal{G}_1 x  = y   \} } \Pr( x ) .  
\end{equation}
For our proof, it is useful to work with unnormalized probabilities that we write as
\begin{equation}
    P_l( y ) =   \sum_{ \{ x :  \mathcal{G}_0 x = 0 , \mathcal{G}_1 x = y   \} } P_{l-1}( x ) ,  
\end{equation}
where we will renormalise later. Herein, we focus on the smallest weight errors that can lead to a particular $y$.  The no error case ($y=0$) occurs when there the initial states had no errors, so
\begin{equation}
    P_l( 0 ) \simeq   (1-\epsilon)^{m_l} ,  
\end{equation}
where $m_l$ counts the total number of raw qubits \textcolor{black}{needed} for $l$ rounds of distillation, namely $m_l=\prod_{j=1, \ldots l} n_j$.  For protocols quadratically suppressing noise, $2^l$ is the smallest number of errors that can evade detection.  Therefore, we take the approximation
\begin{equation}
    P_l( y ) \simeq  C_l(y) \epsilon^{2^l}(1-\epsilon)^{m_l-2^l} ,  
\end{equation}
where $C_l(y)$ counts the number of different weight $2^l$ errors that lead to output $y$.  For one round of distillation this is simply
\begin{equation}
\label{Cend}
    C_1(y) = \ff (y).
\end{equation}
Recall that $\ff (y)$ was defined back in Def.~\ref{distillationFunction} as a function that counts precisely the number of weight 2 \textcolor{black}{errors} that lead to a particular output.  

Finding $C_l(y)$ for higher levels ($l>1$) is more involved.  The relation between input and output error strings is $y=\mathcal{G}_1x$ (assuming $\mathcal{G}_0x=0$), so $x$ vectors of weight 2 can result into a variety of weights $y$ vectors, potentially increasing the weight, and so some care is needed. 

Consider a module (on some $l>1$ level) where some of the incoming branches contain errors.  The probability of two branches $a$ and $b$ containing errors is
\begin{align}
\label{twoBranch}
   &P_{l-1}(x^{(a)})  P_{l-1}(x^{(b)}) P_{l-1}(0)^{n_l-2} \\ \nonumber & \simeq  C_{l-1}(x^{(a)}) C_{l-1}(x^{(b)}) \epsilon^{2^l}(1-\epsilon)^{m_l-2^l}.
\end{align}
\textcolor{black}{These} errors may contribute to undetected errors leaving the module. \textcolor{black}{Fewer or more branch failures} do not provide leading order contributions.  Consider when only 1 branch contains any errors.  After this branch is split up and \textcolor{black}{fed} into different blocks, each $n_l \rightarrow k_l$ block can contain at most 1 error, and so it will be detected. If $t>2$ branches contain an error, the probability will be weighted by $\epsilon^{t*2^{(l-1)}}$, which for small $\epsilon$ is a rare process compared to two failed branches. 

Furthermore, with two erroneous branches $a$ and $b$, we can further deduce that either $x^{(a)}=x^{(b)}$ or the error will be detected.  To see this, we begin by noting that errors will only be undetected if $(G_0 \otimes \id_N)x=0 \pmod{2}$, and we have  
\begin{equation}    
\label{firewall}
    (G_0 \otimes \id_N)x= (G_0 \delta^{(a)}) \otimes x^{(a)} + (G_0 \delta^{(b)}) \otimes x^{(b)} .
\end{equation}
Both $G_0 \delta^{(a)}$ and $G_0 \delta^{(b)}$ are columns of the $G_0$ and so are nonzero.  Since no terms are zero, it can only vanish $\mod 2$ if both $G_0 \delta^{(a)}=G_0 \delta^{(b)}$ and $x^{(a)} = x^{(b)}$.  This is a central pivot of the proof, so we will give a second explanation of what is happening here.  Note that if $x^{(a)} \neq x^{(b)}$, then w.l.o.g $x^{(a)}$ had an error in at least 1 location that is absent from $x^{(b)}$.  If the \textcolor{black}{error's} location is $t$, then the $t^{\mathrm{th}}$ distillation block will receive an input with only 1 error and must detect it.  We conclude that the dominate source of  errors is from the identical failure of pairs of branches. 

To simplify notation, let $u=\delta^{(a)}+\delta^{(b)}$ and $f=x^{(a)}=x^{(b)}$, so we have the more concise expression $x=u \otimes f$ for relevant errors, with $u$ being weight 2. Above we noted that an undetected error must satisfy $G_0 \delta^{(a)}=G_0 \delta^{(b)}$, which is equivalent to $G_0 u=0 \pmod{2}$. The output from such an error is $y=(G_{1}\otimes \id_N)(u \otimes f)=(G_1 u) \otimes f$. Therefore, we can now deduce that
\begin{align}
\label{Pl}
 P_l(v \otimes f ) & = \sum_{\{u : G_0 u = 0 , G_1 u=v \}} P_{l-1}( f )^2 P_{l-1}( 0 )^{n_l-2}. \nonumber
\end{align}
by counting over all $u$ that lead to the same $v$.  Therefore,
\begin{align}
 C_l(v \otimes f ) & = \sum_{\{u : G_0 u = 0 , G_1 u=v, |u|=2 \}} C_{l-1}( f )^2. \nonumber
\end{align}
The summation is over weight 2 vectors $u$, but the terms are independent of $u$, and so we find that 
\begin{align}
 C_l( y ) = C_l(v \otimes f )  = \eta_{l}(v) C_{l-1}( f )^2, \nonumber
\end{align}
where we have again used the $\eta$ notation introduced in Def.~\ref{distillationFunction}.  For two rounds of distillation, $l=2$, and so $C_{l-1}=C_1$ and we can end the recursion by using Eq.~\ref{Cend}, so that
\begin{equation}
 C_2( y ) = C_2(v \otimes f )  = \eta_{2}(v) \eta_{1}( f )^2, 
\end{equation}
However, if we have more than 2 rounds, notice that the relevant $f$ will again have the form $f=v' \otimes f'$ so that
\begin{align}
 C_l( y ) & = \eta_{l}(v) \cdot [C_{l-1}( v' \otimes f' )]^2,\\ \nonumber
   & = \eta_{l}(v) \cdot [ \eta_{l-1}(v') C_{l-2}(f')^2  ]^2, \\ \nonumber
      & = \eta_{l}(v) \cdot [  \eta_{l-1}(v')]^2 \cdot [C_{l-2}(f')]^4, 
\end{align}
and so on, until we reach $C_{1}$ and can use (see Eq.~\ref{Cend}) to end the recursion. 

From $C_l( y )$ we have a good leading order approximation of the probability of an error $P_l(y)$.  The total (unnormalized) probability of an error is
\begin{align}
    B_l &= \sum_{y \neq 0} P_{l}(y) \\ \nonumber
    &= \epsilon^{2^l}(1-\epsilon)^{m_l-2^l} \sum_{y \neq 0} C_{l}(y) \\ \nonumber
    &= \epsilon^{2^l}(1-\epsilon)^{m_l-2^l} C_{l}
\end{align}
where we have used the shorthand
\begin{align}
    C_l &= \sum_{y \neq 0}  C_l( y ),  \\ \nonumber
    &= \prod_{j=1}^{l} \left( \sum_v \eta_{j}(v)^{2^{l-j}} \right).
\end{align}
Equating also $A_l=P_l(0)=(1-\epsilon)^{m_{l}}$, we find we have reached the expressions of Eqs~\ref{ABCexpressions}.  To renormalise the error probability $B_l$, we simply divide through by the total probability $A_l+B_l$, yielding one equation of Thm.~\ref{DistFuncThm}.  The denominator $A_l+B_l$ represents that success probability not just of a single module succeeds, but that all events feeding into that module also succeed.  We are actually interested in the success probability conditioned on previous modules being successful, which is $(A_l+B_l)$ divided by $(A_{l-1}+B_{l-1})^{n_l}$.  This divider here comes from the unconditional success probability of an $(l-1)$-level module,  $(A_{l-1}+B_{l-1})$ and that $n_l$ of these feed into an $l$-level module.  This completes the proof of Thm.~\ref{DistFuncThm}.

\section{Numerical Simulations}
\label{simulations}

\begin{figure*}

\includegraphics{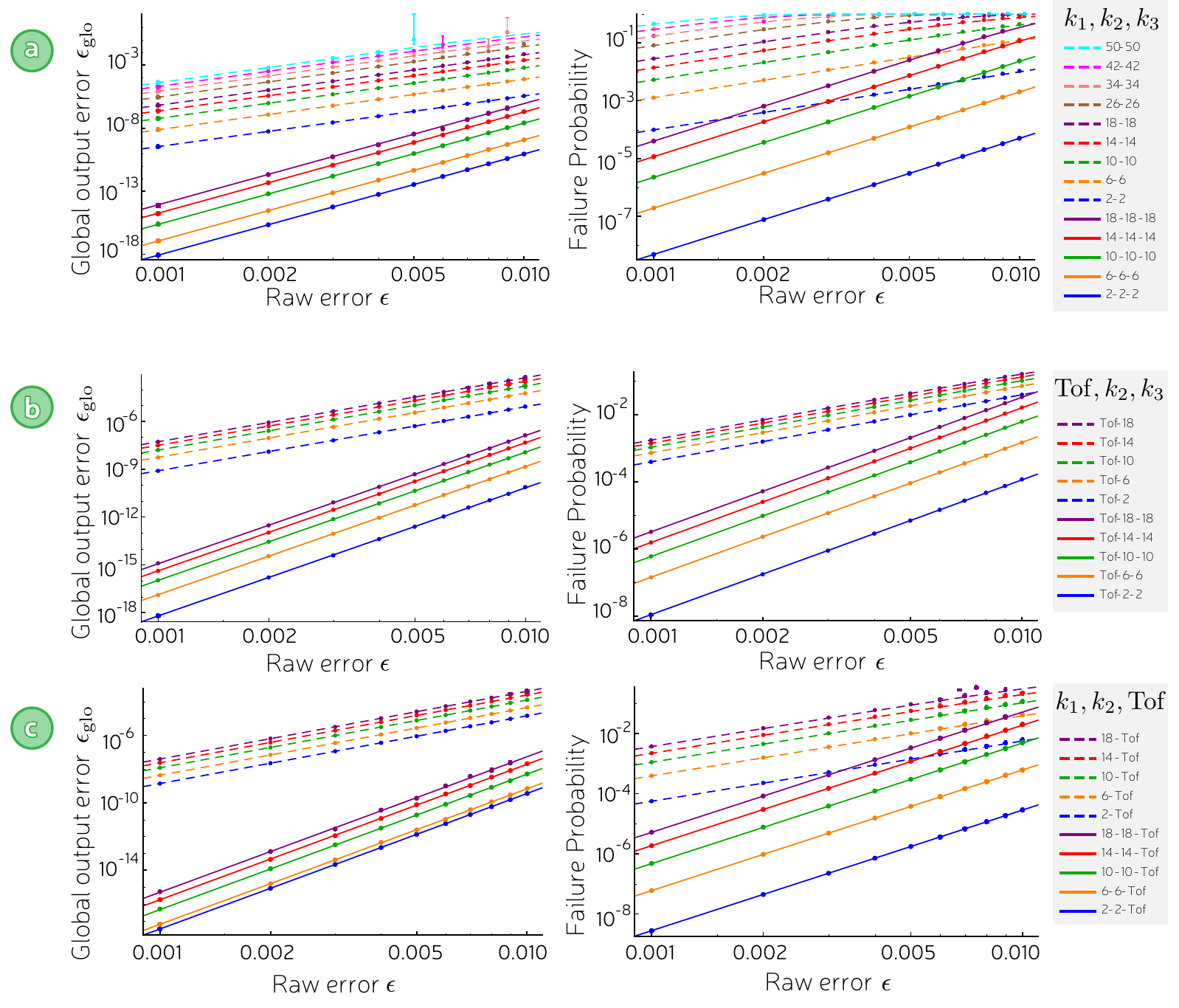} 

\caption{Numeric vs analytic. (a) Bravyi-Haah block protocols with two and three rounds of distillation. (b) Toffoli protocol followed by one or two rounds of Bravyi-Haah. (c) One or two rounds of Bravyi-Haah followed by the Toffoli protocol. The close correspondence of our analytic treatment of the factory and numerical simulations is demonstrated for two and three rounds of the magic state distillation with module checking. Points represent simulation data, while the lines are the corresponding analytic \textcolor{black}{estimates} as determined by Thm.~\ref{DistFuncThm}. The protocols are labelled by the $k$ value of each round, which is set to be equal for each round of distillation. The global error, the probability of a logical error anywhere in the output of the factory is shown. For a ``2-2-2" factory this \textcolor{black}{is} the probability of an error anywhere in the 8 output magic states, for a ``26-26-26'' this would correspond to an error anywhere in the 17,576 magic states that \textcolor{black}{this} factory produces as its output.}
\label{simulationDataA}
\end{figure*}
\subsection{Brute Force method}

In simulating a magic state factory it quickly becomes apparent that a brute force method of simulation is inadequate. The ``brute force'' method simply involves randomly generating a boolean string of length $\prod_{l}n_l$ to describe the input to the magic state factory and performing calculations of the stabilizers and logical output of each module of the factory. The explicit procedure for a three round factory is given in Alg.~\ref{alg::bf}. For each module in round one, a random input is generated and \textcolor{black}{tested} until the stabilizer checks for the module are passed. The logical output of each of these successful modules is saved until enough have been generated to feed into round two. These outputs are shuffled according to Equation~\ref{firewall} before entering round two. Here the module checking is performed again. If all the module checks are passed, then we again proceed by feeding the logical output of round 2 to round 3, after shuffling. Having reached round 3 we then determine the characteristics of this module by recording where the module check fails, and if it succeeds whether the output of the module contains a logical (undetected error).

Clearly a large number of iterations of this protocol are required to obtain reliable statistics for the performance of the factory. For example, our analytic treatment estimates that with a raw magic state error rate $\epsilon=0.001$ that the rate of undetected logical error of a round 3 module is $\sim10^{-21}$. As such, successfully simulating this by brute force would require $\gg10^{21} $ attempts at the algorithm to be made, not just to build statistics but to ensure that enough instances of the third round module are generated in the first place. We find that \textcolor{black}{the simulation of }two round factories by brute force is achievable for $2<k<50$, but to simulate three rounds a different method is required.

\subsection{Rare Events method}

An undetected error in the factory's output after three rounds of distillation is a rare event. To simulate these and gain adequate statistics we must use a method in which we as far \textcolor{black}{as} possible eliminate the simulations of input error configurations that we know cannot lead to a logical error. Our chosen method of doing this proceeds as follows. We know that a module only has a chance of failing if at least two of the branches entering that module contain an error.   For a module in the third round, we focus on cases where at least 2 of its input branches contain errors. We do this by a process that can be called preselection in contrast to postselection.  In postselection, we sample from a distribution and reject instances that do not meet a certain criteria.  This is not feasible when the criteria (here having at least 2 corrupt branches) is rare, and so we instead construct a new probability distribution conditioned on the criteria being meet.  The first step of the algorithm therefore is to first decide how many error containing branches will enter this module, given that this number is $\geq 2$, based on the statistics already gathered for the rate of errors in the output of round 2 modules.  Later we analytically adjust for this process of preselection.

\begin{figure}
\includegraphics{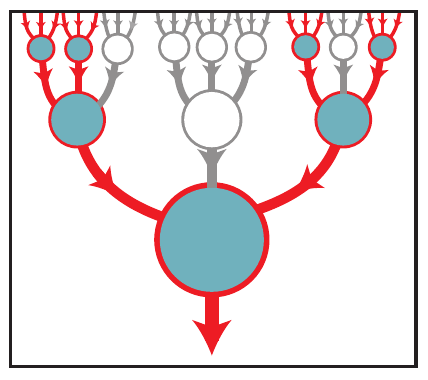}
\caption{Rare events method. When simulating the fictional factory of Fig.~\ref{FIGtree} with our `Rare Events' method, only the red highlighted parts of the factory are actively simulated, i.e. require computation of the stabilizer outcomes and the logical state of the outputs.}
\label{rareEventsFig}
\end{figure}

For each of the `corrupt' branches entering round three, we know that these must originate from a round 2 module which itself had 2 or more corrupt branches entering it. These branches again would have originated in a round 1 module (equivalently a block) which had at least two errors fed to it. We can thus greatly reduce the size of the simulation and the number of iterations required by simulating only on a subsection of the factory where these errors have occurred, see Fig.~\ref{rareEventsFig}. The probability of an undetected error after the first round was determined analytically for Bravyi-Haah by explicitly calculating the weight enumerator \begin{equation}
\begin{split}
W(k,\epsilon)&=2 \sum_{m \text{ is odd}}^{k} (1-2\epsilon)^{3m+4}+\sum_{m=0}^{k/2} (1-2\epsilon)^{3(2m)}\\
&+6\sum_{m =0}^{k} (1-2\epsilon)^{2k-m+4}+\sum_{m=0}^{k/2} (1-2\epsilon)^{3(2m)+8}
\end{split}
\end{equation}
which allows analytic calculation of the global fidelity of the output of a single round of distillation. The corresponding result for the Toffoli protocol is given in Ref~\cite{eastin13}.

Having chosen the number of failed branches that need to enter round three we attempt to generate these failed branches. We thus simulate the reduced factory, as described above, discarding and repeating each module not only if the stabilizer module check has failed, but also when the check is passed and the output does not contain a logical error. Using this method we eliminate the simulation of a vast number of the possible input error strings, while holding smaller boolean vectors in memory at all but the last step, the full simulation of the module in round 3. Some pseudo-code is presented in Alg.~\ref{alg::re}.

This method of simulation allows us to calculate the correlated error rate as follows.

A key number is the probability $p_{\mathrm{num}}$ that two or more branches leaving round 2 of the factory contain an error.

\begin{equation}
p_{\mathrm{num}}=1-(1-\epsilon^{(2)}_\mathrm{glo})^{n_3} -n_3\epsilon^{(2)}_\mathrm{glo} (1-\epsilon^{(2)}_\mathrm{glo})^{n_3-1}
\end{equation}
where $\epsilon^{(2)}_\mathrm{glo}$ is the probability that a branch exiting round 2 contains an undetected error, as determined by numerical simulations of a two round factory. 

Using this allows us to determine the success probability and the global error in the output of the round 3 module.

\begin{equation}
p_{\mathrm{suc}}^{(3)}=(1-\epsilon^{(2)}_\mathrm{glo})^{n_3} + p_{\mathrm{num}} (a_3+ b_3) 
\end{equation}

and

\begin{equation}
\epsilon_\mathrm{glo}^{(3)}=\frac{b_3 \  p_{\mathrm{num}} }{p_{\mathrm{suc}}} 
\end{equation}

where $b_3 = \frac{\#\{``\text{ERROR}"\}}{\#\{``\text{SUCCESS}"\}+\#\{``\text{FAIL}"\}+\#\{``\text{ERROR}"\}}$ is the estimate of the probability of a logical error in the output branch of round 3 from output of the simulation; and $a_3 = \frac{\#\{``SUCCESS"\}}{\#\{``\text{SUCCESS}"\}+\#\{``\text{FAIL}"\}+\#\{``\text{ERROR}"\}}$ 

\begin{algorithm}
\begin{algorithmic}[1]
\STATE{select protocol: $\{k_1,k_2,k_\text{Toff}\}$}
\STATE{generate number of modules in each round: $\{M_1,M_2,M_\text{Toff}\}$}
\COMMENT{Round One}
\FOR{$i <M_1$}
	\STATE{randomly generate binary string $v$ length $n_1$}
	\STATE{measure stabilizers $G_0(k_1).v$}
	\IF{stabilizers failed} return to line 3
	\ENDIF{}
	\STATE calculate logical output of module $v = G_1(k_1).v$
	\STATE Append $v$ to list of logical outputs $V$
\ENDFOR 
\STATE shuffle output of round 1 to firewall correlations $V\rightarrow V'$
\COMMENT{Round Two}
\FOR{$j<M_2$}
        \FOR{each block $ii$ in a round 2 module (there are $k_1$)}
    		\STATE{take $(ii\times j)^{\text{th}}$ string of length $n_2$ from $V'$: $v'$}
		\STATE{measure stabilizers $G_0(k_2).v'$}
		\IF{stabilizers failed} return to line 1
		\ENDIF{}
		\STATE calculate logical output of module $w=G_1(k_2).v'$
		\STATE Append $w$ to list of logical outputs $W$
	\ENDFOR 
\ENDFOR
\STATE shuffle output of round 2 to firewall correlations $W\rightarrow W'$
\COMMENT{Round Three}
\FOR{$l<k_1k_2$}
	\STATE{measure stabilizers $G_0(k_3).w'$}
	\IF{stabilizers failed} \textbf{return} FAIL
	\ENDIF{ stabilizers passed}
	\STATE calculate logical output of module $G_1(k_3).w'$
	\STATE Search for logical error in output
	\IF{logical error found} \textbf{return} ERROR
	\ELSE{logical error not found } \textbf{return} SUCCESS
	\ENDIF
\ENDFOR 
\end{algorithmic}
\caption{Brute force simulation algorithm}
\label{alg::bf}
\end{algorithm}

\begin{algorithm}
\begin{algorithmic}[1]
\STATE{select protocol: $\{k_1,k_2,k_\text{Toff}\}$}
\STATE{generate number of modules in each round: $\{M_1,M_2,M_\text{Toff}\}$}
\STATE{generate number of corrupt modules in round 2: $N_2$}
\COMMENT{Round One}
\FOR{$i < N_2$}
	\STATE{generate number of corrupt round 1 modules $N_1$}
	\FOR{$j < N_1$}
	\STATE{randomly generate binary string $v$ length $n_1$}
	\STATE{measure stabilizers $G_0(k_1).v$}
		\IF{stabilizers failed} return to line 6
		\ENDIF{}
		\STATE calculate logical output of module $G_1.v$
		\IF{there is a logical error} Append $v$ to list of logical outputs $V$
		\ELSE{ return to line 6}
		\ENDIF{}
	\ENDFOR
		
	\STATE Pad $V$ with 0s so it is length $k_1n_2$
	\STATE shuffle output of corrupt module to firewall correlations $\ ... \  V\rightarrow V'$
	
	\COMMENT{Round Two}
        \FOR{each block $ii$ in a round 2 module (there are $k_1$)}
    		\STATE{take $(ii)^{\text{th}}$ string of length $n_2$ from $V'$: $v'$}
		\STATE{measure stabilizers $G_0(k_2).v'$}
		\IF{stabilizers failed} return to line 6
		\ENDIF{}
		\STATE calculate logical output of module $w=G_1(k_2).v'$
		\IF{there is a logical error} Append $w$ to list $W$
		\ELSE{ return to line 6}
		\ENDIF{}
	\ENDFOR 
\ENDFOR
\COMMENT{ Round Three }
\STATE Pad $W$ with $0$s so it is length $k_1k_2n_3$
\STATE shuffle output of round 2 to firewall correlations $W\rightarrow W'$
\FOR{$l<k_1k_2$}
	\STATE{measure stabilizers $G_0(k_3).w'$}
	\IF{stabilizers failed} \textbf{return} FAIL
	\ENDIF{ stabilizers passed}
	\STATE calculate logical output of module $G_1(k_3).w'$
	\STATE Search for logical error in output
	\IF{logical error found} \textbf{return} ERROR
	\ELSE{logical error not found } \textbf{return} SUCCESS
	\ENDIF
\ENDFOR 
\end{algorithmic}
\caption{Rare events simulation algorithm}
\label{alg::re}
\end{algorithm}

We limit our simulations to a limited set of k values, with $k_i=k$ for each simulation. This is an arbitrary choice and simplifies the comparison made in Fig.~\ref{simulationDataA}. For three rounds the simulations are limited to low k values, as these can be simulated in a reasonable timeframe and adequate statistics gathered to infer the output error rate. 
\clearpage
\end{document}